\documentclass[journal,vlined,ruled]{IEEEtran}

\usepackage[latin9]{inputenc}
\usepackage{color}
\usepackage{verbatim}
\usepackage{enumitem}
\usepackage{algorithm2e}
\usepackage{amsmath}
\usepackage{amsthm}
\usepackage{amssymb}
\usepackage{graphicx}
\usepackage{wasysym}

\makeatletter

\DeclareTextSymbolDefault{\textquotedbl}{T1}
\providecommand{\tabularnewline}{\\}

\let\oldforeign@language\foreign@language
\DeclareRobustCommand{\foreign@language}[1]{%
  \lowercase{\oldforeign@language{#1}}}
\theoremstyle{plain}
\newtheorem{thm}{\protect\theoremname}
\theoremstyle{plain}
\newtheorem{lem}[thm]{\protect\lemmaname}
\theoremstyle{definition}
\newtheorem{defn}[thm]{\protect\definitionname}
\theoremstyle{plain}
\newtheorem{prop}[thm]{\protect\propositionname}
\newlist{casenv}{enumerate}{4}
\setlist[casenv]{leftmargin=*,align=left,widest={iiii}}
\setlist[casenv,1]{label={{\itshape\ \casename} \arabic*.},ref=\arabic*}
\setlist[casenv,2]{label={{\itshape\ \casename} \roman*.},ref=\roman*}
\setlist[casenv,3]{label={{\itshape\ \casename\ \alph*.}},ref=\alph*}
\setlist[casenv,4]{label={{\itshape\ \casename} \arabic*.},ref=\arabic*}

\usepackage{amsfonts,times,amsmath,amssymb,amsthm}

\def\app#1#2{%
  \mathrel{%
    \setbox0=\hbox{$#1\sim$}%
    \setbox2=\hbox{%
      \rlap{\hbox{$#1\propto$}}%
      \lower1.1\ht0\box0%
    }%
    \raise0.25\ht2\box2%
  }%
}
\def\approxprop{\mathpalette\app\relax}

\@ifundefined{showcaptionsetup}{}{%
 \PassOptionsToPackage{caption=false}{subfig}}
\usepackage{subfig}
\makeatother

\providecommand{\casename}{Case}
\providecommand{\definitionname}{Definition}
\providecommand{\lemmaname}{Lemma}
\providecommand{\propositionname}{Proposition}
\providecommand{\theoremname}{Theorem}

\begin{document}
\markboth{}{}
\title{Multi-Scan Multi-Sensor Multi-Object State Estimation }
\author{Diluka Moratuwage, Ba-Ngu Vo, Ba-Tuong Vo and Changbeom Shim \thanks{Acknowledgment: This work is supported by the Australian Research
Council under Linkage project LP200301507 and Future Fellowship FT210100506.} \thanks{The authors are with the Department of Electrical and Computer Engineering,
Curtin University, Bentley, WA 6102, Australia (email: \{diluka.moratuwage,
ba-ngu.vo, ba-tuong.vo, changbeom.shim\}@curtin.edu.au). 

This paper has supplementary downloadable material available at http://ieeexplore.ieee.org, provided by the author. This material includes proofs of the lemmas and propositions presented in the paper. This material is 280KB in size.}}

\markboth{PREPRINT: IEEE TRANSACTIONS ON SIGNAL PROCESSING, VOL. 70, PP. 5429-5442, 2022.}%
{}

\maketitle
\begin{abstract}
If computational tractability were not an issue, multi-object estimation
should integrate all measurements from multiple sensors across multiple
scans. In this article, we propose an efficient numerical solution
to the multi-scan multi-sensor multi-object estimation problem by
computing the (labeled) multi-sensor multi-object posterior density.
Minimizing the $L_{1}$-norm error from the exact posterior density
requires solving large-scale multi-dimensional assignment problems
that are NP-hard. An efficient multi-dimensional assignment algorithm
is developed based on Gibbs sampling, together with convergence analysis.
The resulting multi-scan multi-sensor multi-object estimation algorithm
can be applied either offline in one batch or recursively. The efficacy
of the algorithm is demonstrated using numerical experiments with
a simulated dataset.
\end{abstract}


\section{Introduction\label{sec:Intro}}

Instead of the state or trajectory of a single object, multi-object
state estimation is concerned with the joint estimation of the number
of objects and their trajectories \cite{Mahler2007a}, \cite{Blackman1999},
\cite{Mahler2014}. This problem has a wide range of application areas
from aerospace \cite{Bar-Shalom2011}, computer vision\textcolor{black}{{}
\cite{Szeliski2010}, \cite{Ong2022}, \cite{Ishtiaq2022},} to cell
biology \cite{Meijering2012}\cite{Nguyen2021}. Multi-object state
estimation is challenging because we have to address the unknown and
time-varying number of objects, false negatives/positives, and data
association uncertainty, which altogether incur a combinatorial complexity.
The three main approaches to multi-object estimation in the literature
are Multiple Hypothesis Tracking (MHT) \cite{Blackman1999}, Joint
Probabilistic Data Association (JPDA) \cite{Bar-Shalom1988}, and
Random Finite Set (RFS) \cite{Mahler2007a,Mahler2014}. Apart from
the number of objects and their trajectories, the RFS approach also
provides statistical characterization of the entire ensemble of objects
\cite{Vo2019b}.

In general, employing multiple sensors in state estimation enhances
detection capability and spatial coverage, reliability, observability,
and reduces uncertainty \cite{Liggins2008}, \cite{Chen2014}. There
are many applications that fundamentally require multiple sensors
because a single sensor is simply not adequate, see e.g., \cite{Thrun2006,Groves2013}.
Moreover, given the proliferation of inexpensive sensors, effective
multiple sensor solutions are imperative for exploiting their intended
capabilities. The benefits of multiple sensors are even more significant
in multi-object estimation, where there is inherent additional uncertainty. 

In multi-object estimation, the benefits of multiple sensors come
with additional challenges due to the NP-hard multi-dimensional
data association problem that arises from the matching of objects
to measurements across multiple sensors \cite{Mahler2014}, \cite{Chen2014},
\cite{Bar-Shalom1995}, \cite{Reynen2019}. Several approximate multi-sensor
multi-object estimation solutions have been developed. Centralized
architecture solutions include the multi-sensor Probability Hypothesis
Density (PHD) and Cardinalized PHD (CPHD) filters \cite{Mahler2014},
\cite{Pham2007} multi-sensor JPDA filter \cite{Meyer2017}, multi-Sensor
multi-Bernoulli filter \cite{Saucan2017}, and multi-sensor Generalized
Labeled Multi-Bernoulli (GLMB) filter \cite{Vo2019a}. The latter
incurs a linear complexity in the total number of measurements across
the sensors, without significantly compromising optimality. Decentralized
solutions have also been developed for the PHD, CPHD filters \cite{Uney2013},
\cite{Battistelli2015}, \cite{Battistelli2013} multi-Bernoulli filter
\cite{Guldogan2014}, \cite{Jian2016}, \cite{Wang2017}, LMB and
marginalized GLMB filters \cite{Fantacci2018}.

Whereas filtering only considers the current timestep, the multi-scan
approach considers the history of the multi-object states over a time
window, thereby allowing the correction of previous errors \cite{Drummond1993},
\cite{Blackman1999}, \cite{Vu2014}. MHT forms hypotheses by associating
measurements to tracks across a time window \cite{Reid1979}, and
delays making estimates until further information becomes available.
In \cite{Mahalanabis1990}, a fixed lag smoother was integrated to
the JPDA filter \cite{Bar-Shalom1988}, where it was shown that a
significant improvement in tracking accuracy over filtering can be
achieved with small window of sizes two and three. The first RFS multi-scan
multi-object filter was developed in \cite{Vu2014}, where all statistical
information pertinent to the multi-object system over the time window
is encapsulated in the multi-object posterior, including statistics
on the ensemble of trajectories. Computing this posterior, however,
also requires solving NP-hard multi-dimensional ranked assignment
problems. Recently, an efficient numerical algorithm with polynomial
complexity was developed for propagating the so-called (multi-scan)
GLMB posterior, where multi-dimensional ranked assignment problems
are solved using Gibbs sampling \cite{Vo2019b}. 

Ideally, if computational tractability were not an issue, to inherit
all the benefits from both multi-sensor and multi-scan solutions we
should make use of measurements from all sensors across multiple scans.
However, keeping in mind that the multi-scan problem itself requires
solving NP-hard multi-dimensional ranked assignment problems, and
ditto for the multi-sensor problem, the joint multi-scan multi-sensor
problem is far more computationally demanding. Consequently, research
on multi-scan multi-sensor multi-object estimation is very limited.
To the best of our knowledge, only a two-scan two-sensor demonstration
with four objects, using an Interacting Multiple Model (IMM) JPDA filter,
has been reported \cite{Sumedh2005}.

This article addresses multi-scan multi-sensor multi-object estimation
via the labeled RFS formulation, which translates to computing a labeled
multi-object posterior consisting of an intractably large weighted
sum of set functions. Approximating this so-called GLMB posterior
by retaining the highest weighted terms minimizes the $L_{1}$-norm
approximation error \cite{Vo2019b}, and requires solving far more
challenging multi-dimensional assignment problems than those for multi-scan
(with single-sensor) or multi-sensor (with single-scan). Indeed, the
posterior truncation problem for \textit{V} sensors across \textit{K}
scans is a (\textit{KV})-dimensional assignment problem, and existing
techniques are not adequate for \textit{K} and \textit{V} greater
than two, even with only four objects. 

We propose an efficient multi-dimensional assignment solution with
polynomial complexity using Gibbs sampling, together with convergence
analysis. This solution generalizes those for multi-sensor (with single-scan)
and multi-scan (with single-sensor) in \cite{Vo2019a}, \cite{Vo2019b}.
The resultant multi-object estimation algorithm can be used offline
in one batch or recursively at each measurement update, i.e., smoothing-while-filtering.
The efficacy and utility of the proposed solution are demonstrated
in numerical experiments. For the purposes of establishing the theoretical
foundation and scalability for the algorithm, we focus on the (labeled)
multi-object posterior, which involves a growing window of scans.
While such a growing window results in a complexity per time step
that grows with time, the algorithm can be easily adapted to a fixed-length
moving window, so that the complexity per time step is fixed. Note
that filtering is a special case of the moving window approach with
a window length of one. A formulation that produces trajectories with
a fixed complexity per time step is only possible with labels, see
e.g., \cite{Mahler2019}, \cite{Mahler 2022} for further discussion.
Without labels, the only way to obtain trajectories is by using a
growing window which incur a complexity per time step that grows exponentially
with time. Such an approach is impractical even for the simplest case
of single object state estimation with a single sensor \cite{Sarkka2013}. 

The remainder of this article is organized as follows. Section \ref{sec:Backgnd}
summarizes relevant concepts in multi-object estimation, and Section
\ref{sec:Comp_MSMS_GLMB_Posts} presents the implementation details
of the multi-sensor GLMB posterior recursion. Section \ref{sec:Num_Exps}
presents the numerical studies, and Section \ref{sec:Conclusion}
concludes the paper. Mathematical proofs are given in the supplementary
materials.

\section{Background \label{sec:Backgnd}}

As per the convention in \cite{Vo2019b}, the symbols and notations
used throughout the article are summarized in Table \ref{tab:Notations}.
\begin{table}
\caption{\label{tab:Notations}Notations}

\centering{}%
\begin{tabular}{|c|l|}
\hline 
Notation & Description\tabularnewline
\hline 
$X_{m:n}$ & $X_{m},X_{m+1},\ldots,X_{n}$\tabularnewline
$1_{S}(\cdot)$ & Indicator function for a given set $S$\tabularnewline
$\delta_{X}[Y]$ & Kroneker-$\delta$, $\delta_{X}[Y]=1$ if $X=Y$, 0 otherwise\tabularnewline
$\mathcal{F}(S)$ & Class of all finite subsets of a given set $S$\tabularnewline
$\langle f,g\rangle$ & Inner product, $\int f(x)g(x)dx$ of two functions $f$ and $g$\tabularnewline
$|X|$ & Cardinality of a set $X$\tabularnewline
$h^{X}$ & Multi-object exponential, $\prod_{x\in X}h(x)$, with $h^{\emptyset}=1$\tabularnewline
$1^{V}$ & V-dimensional vector $(1,...,1)$\tabularnewline
$\mathcal{L}(\boldsymbol{x})$ & Label of a (labeled) state $\boldsymbol{x}$, $\mathcal{L}((x,\ell))=\ell$\tabularnewline
$\mathcal{L}(\boldsymbol{X})$ & Labels of a (labeled) set $\boldsymbol{X}$, $\{\mathcal{L}(\boldsymbol{x}):\boldsymbol{x}\in\boldsymbol{X}\}$\tabularnewline
$\Delta(\boldsymbol{X})$ & Distinct label indicator $\delta_{|\boldsymbol{X}|}(|\mathcal{L}(\boldsymbol{X})|)$\tabularnewline
\hline 
\end{tabular}
\end{table}
\begin{figure}
\includegraphics[scale=0.5]{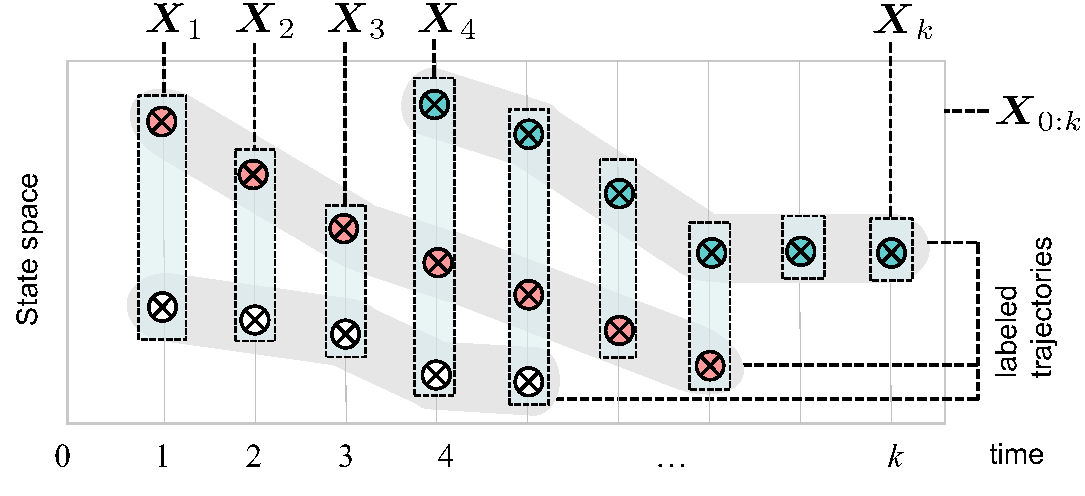}

\caption{\label{fig:Labeled-multi-object-trajectory}Labeled multi-object trajectory
and history. The two objects born at time 1 are given labels (1,1)
and (1,2), colored red and white. The object born at time 4 is given
label (4,1), colored blue. The multi-object history $\boldsymbol{X}_{0:k}$
(note $\boldsymbol{X}_{0}=\emptyset$) is represented by two equivalent
groupings according to: (a) time (vertical strips, i.e., multi-object
states) or; (b) labels (strips containing states of the same color,
i.e., trajectories).}
\end{figure}

\subsection{Multi-object State and Multi-object Trajectory\label{ss:Multi-object-States-and}}

The labeled state of an object is represented by $\boldsymbol{x}=(x,\ell)$,
where the vector $x\in\mathbb{X}$ is its kinematic state, and $\ell=(s,\iota)$
is its (unique) label with $s$ representing the time of birth and
$\iota$ is an index to distinguish objects born at the same time
\cite{Vo2013}. Let $\mathbb{B}_{s}$ denote the label space of objects
born at time $s$, then the space of all labels up to time $k$ is
the disjoint union $\mathbb{L}_{k}=\uplus_{s=0}^{k}\mathbb{B}_{s}$
(equivalently $\mathbb{L}_{k}=\mathbb{L}_{k-1}\uplus\mathbb{B}_{k}$).
A sequence of consecutive labeled states (say, from time $s$ to $t$)
\begin{equation}
\boldsymbol{\tau}=[(x_{s},\ell),(x_{s+1},\ell),\ldots,(x_{t},\ell)],\label{eq:trajectory}
\end{equation}
with kinematic states $x_{s},x_{s+1},...,x_{t}\in\mathbb{X}$, and
label $\ell=(s,\iota)$, is called a \textit{trajectory}. The labeled
state of trajectory $\boldsymbol{\tau}$ at time $i$ is denoted by
$\boldsymbol{\tau}(i)$. 

At time $i$, a \textit{labeled multi-object state} $\boldsymbol{X}_{i}$
is a finite subset of $\mathbb{X}\times\mathbb{L}_{i}$ with distinct
labels. Let $\mathcal{L}$ denote the projection defined by $\mathcal{L}((x,\ell))=\ell$,
and $\mathcal{L}(\boldsymbol{X}_{i})$ denote the set of labels of
$\boldsymbol{X}_{i}$. Since a valid labeled multi-object state $\boldsymbol{X}_{i}$
must have distinct labels, we require the \textit{distinct label indicator
}$\Delta(\boldsymbol{X}_{i})\triangleq\delta_{|\boldsymbol{X}_{i}|}[|\mathcal{L}(\boldsymbol{X}_{i})|]$
to be 1. Given a set $S$ of trajectories with distinct labels, the
labeled multi-object state at time $i$ is $\boldsymbol{X}_{i}=\{\boldsymbol{\tau}(i):\boldsymbol{\tau}\in S\}$,
see Fig. \ref{fig:Labeled-multi-object-trajectory} for illustration.

Given a sequence $\boldsymbol{X}_{j:k}$ of labeled multi-object states
(from a set of labeled trajectories) over the interval $\{j:k\}$,
the trajectory of (the object with) label $\ell\in\cup_{i=j}^{k}\mathcal{L}(\boldsymbol{X}_{i})$
is
\begin{equation}
\boldsymbol{x}_{s(\ell):t(\ell)}^{(\ell)}\triangleq[(x_{s(l)}^{(\ell)},\ell),\ldots,(x_{t(l)}^{(\ell)},\ell)],\label{eq:trajectory_as_state_seq}
\end{equation}
where: $s(\ell)$ and $t(\ell)$ are, respectively, the earliest and
latest times on $\{j:k\}$ such that the label $\ell$ exists; and
$(x_{i}^{(\ell)},\ell)$ denotes the element of $\boldsymbol{X}_{i}$
with label $\ell$ and unlabeled state $x_{i}^{(\ell)}$. Hence, the
sequence $\boldsymbol{X}_{j:k}$ can be equivalently represented by
the trajectories of all labels in $\cup_{i=j}^{k}\mathcal{L}(\boldsymbol{X}_{i})$,
i.e.,
\begin{equation}
\mathbb{\mathbf{{\boldsymbol{X}}_{\mathnormal{{j:k}}}\equiv\big\{\mathnormal{{\boldsymbol{x}_{s(\ell):t(\ell)}^{(\ell)}:\ell\in\cup_{i=j}^{k}\mathcal{L}(\mathbf{{\boldsymbol{X}}}_{i})}}\big\}}}.\label{eq:multi_obj_state_seq}
\end{equation}
This equivalence is illustrated in Fig. \ref{fig:Labeled-multi-object-trajectory}.

Analogous to a single-object system where the state\textit{ trajectory}
is a represented by the state history $x_{0:k}$, in a multi-object
system the \textit{multi-object trajectory} is represented by the
labeled multi-object (state) history $\boldsymbol{X}_{0:k}$. Further,
the equivalence (\ref{eq:multi_obj_state_seq}) enables the following
extension of the multi-object exponential to multiple scans. For any
non-negative integer $n$ and $i_{1}<i_{2}<...<i_{n}$, let $\mathbb{T}_{\{i_{1},i_{2},...,i_{n}\}}\triangleq\big(\mathbb{X}\times\mathbb{L}_{i_{1}}\big)\times....\times\big(\mathbb{X}\times\mathbb{L}_{i_{n}}\big)$,
with $\mathbb{T}_{\emptyset}=\emptyset$. For any function $h:\uplus_{I\subseteq\{j:k\}}\mathbb{T}_{I}\rightarrow[0,\infty)$,
the \textit{multi-scan exponential }\cite{Vo2019b} is defined as
\[
[h]^{\boldsymbol{X}_{j:k}}\triangleq[h]^{\{\boldsymbol{x}_{s(\ell):t(\ell)}^{(\ell)}:\ell\in\mathcal{L}(\boldsymbol{X}_{j:k})\}}=\underset{\ell\in\mathcal{L}(\boldsymbol{X}_{j:k})}{\prod}h(\boldsymbol{x}_{s(\ell):t(\ell)}^{(\ell)}).
\]
Note that when $j=k$, this reduces to the single-scan multi-object
exponential $h^{\boldsymbol{X}_{j}}$ in Table 1.

Hereon, single object states are represented by lower case letters
(i.e., $x$ and $\boldsymbol{x}$), and multi-object states are represented
by upper case letters (i.e., $X$ and $\boldsymbol{X}$), where the
symbols for labeled states and their distributions are bolded (i.e.,
$\boldsymbol{x}$, $\boldsymbol{X}$, $\boldsymbol{\pi}$, etc.) to
distinguish them from unlabeled states.

In Bayesian estimation, the states and measurements are modeled as
random variables. Hence, in a multi-object system, the multi-object
state is modeled as an RFS, and the system model is described by the
multi-object state transition kernel and measurement likelihood function,
presented respectively in Subsections \ref{ss:Multi-Object-Dynamic-Model}
and \ref{ss:Multi-Object-Measurement-Model}. The multi-object posterior
recursion is presented in Subsection \ref{ss:Multi-Object-Bayes-Recursion}.

\subsection{Multi-object Dynamic Model\label{ss:Multi-Object-Dynamic-Model}}

Given the multi-object state $\boldsymbol{X}_{k-1}$ at time $k-1$,
each $\boldsymbol{x}_{k-1}=(x_{k-1},\ell_{k-1})\in\boldsymbol{X}_{k-1}$
either survives with probability $P_{S,k-1}(\boldsymbol{x}_{k-1})$
and moves to a new state $\boldsymbol{x}_{k}=(x_{k},\ell_{k})$ with
transition density $f_{S,k|k-1}(x_{k}|x_{k-1},\ell_{k})\delta_{\ell_{k-1}}[\ell_{k}]$
at time $k$, or dies with probability $Q_{S,k-1}(\boldsymbol{x}_{k-1})=1-P_{S,k-1}(\boldsymbol{x}_{k-1})$.
An object (with label) $\ell_{k}\in\mathbb{B}_{k}$, is either born
at time $k$ with probability $P_{B,k}(\ell_{k})$ and state $x_{k}$
with probability density $f_{B,k}(x_{k},\ell_{k})$, or not born with
probability $Q_{B,k}(\ell_{k})=1-P_{B,k}(\ell_{k})$. The multi-object
state $\boldsymbol{X}_{k}$ at time $k$, is the superposition of
surviving and new birth states. In a standard multi-object dynamic
model, the birth and survival sets are independent of each other,
and each object moves and dies independently of each other. The multi-object
dynamic model is encapsulated in the multi-object transition density
$\boldsymbol{f}_{k|k-1}(\boldsymbol{X}_{k}|\boldsymbol{X}_{k-1})$,
see \cite{Vo2019b,Vo2013} for the actual expressions. 

\subsection{Multi-object Measurement Model\label{ss:Multi-Object-Measurement-Model}}

Consider the multi-object state $\boldsymbol{X}_{k}$ at time $k$,
and $V$ sensors, each produces a set $Z_{k}^{(v)}$ of measurements,
$v\in\{1:V\}$. Each $\boldsymbol{x}\in\boldsymbol{X}_{k}$ is either
detected by sensor $v$ with probability $P_{D,k}^{(v)}(\boldsymbol{x})$
and generates a detection $z\in Z_{k}^{(v)}$ on the observation space
$\mathbb{Z}^{(v)}$, with likelihood $g_{k}^{(v)}(z|\boldsymbol{x})$
or miss-detected with probability $Q_{D,k}^{(v)}(\boldsymbol{x})=1-P_{D,k}^{(v)}(\boldsymbol{x})$.
Sensor $v$ also receives false alarms (clutter), modeled by a Poisson
RFS with intensity function $\kappa_{k}^{\left(v\right)}$ on $\mathbb{Z}^{(v)}$.
The multi-object observation $Z_{k}^{(v)}$ is the superposition of
detections and false alarms, which are assumed independent, conditional
on $\boldsymbol{X}_{k}$.

An \textit{association map} $\gamma_{k}^{(v)}:\mathbb{L}_{k}\rightarrow\{-1:|Z_{k}^{(v)}|\}$,
of sensor $v$, is a positive 1-1 map (i.e., no two distinct labels
are mapped to the same positive value). Here $\gamma_{k}^{(v)}(\ell)>0$
means $\ell$ generates the $\gamma_{k}^{(v)}(\ell)$-th measurement
at sensor $v$, $\gamma_{k}^{(v)}(\ell)=0$ means $\ell$ is misdetected
by sensor $v$, and $\gamma_{k}^{(v)}(\ell)=-1$ means $\ell$ does
not exist. Let $\mathcal{L}(\gamma_{k}^{(v)})\triangleq\{\ell\in\mathbb{L}_{k}:\gamma_{k}^{(v)}(\ell)\geq0\}$
be the set of \textit{live labels}\footnote{This notation is similar to $\mathcal{L}(\boldsymbol{X})$, the labels
of a labeled set, nonetheless the context is clear from the arguments.} of $\gamma_{k}^{(v)}$, and $\Gamma_{k}^{(v)}$ be the space of all
association maps for sensor $v$, then the multi-object likelihood
is \cite{Vo2019a} 
\begin{align}
g_{k}^{(v)}(Z_{k}^{(v)}| & \boldsymbol{X}_{k})\nonumber \\
\propto & \underset{\gamma_{k}^{\left(v\right)}\in\Gamma_{k}^{\left(v\right)}}{\sum}\delta_{\mathcal{L}(\gamma_{k}^{(v)})}[\mathcal{L}(\boldsymbol{X}_{k})][\psi_{k,Z_{k}^{(v)}}^{(v,\gamma_{k}^{(v)}\circ\mathcal{L}(\cdot))}(\cdot)]^{\boldsymbol{X}_{k}}\label{eq:likelihood_func_sensor_v}
\end{align}
where $\gamma_{k}^{(v)}\circ\mathcal{L}(\cdot)=\gamma_{k}^{(v)}(\mathcal{L}(\cdot))$,
and 
\begin{equation}
\psi_{k,\{z_{1:m}\}}^{(v,i)}(\boldsymbol{x})=\begin{cases}
\frac{P_{D,k}^{(v)}(\boldsymbol{x})g_{k}^{(v)}(z_{i}|\boldsymbol{x})}{\kappa_{k}^{(v)}(z_{i})} & i>0\\
Q_{D,k}^{(v)}(\boldsymbol{x}) & i=0
\end{cases}.\label{eq:likelihood_update_func_sensor_v_2}
\end{equation}

The \textit{multi-sensor association map} is defined from the single-sensor
association maps $\gamma_{k}^{(v)}$, $v\in\{1:V\}$ by 
\[
\gamma_{k}=(\gamma_{k}^{(1)},\ldots,\gamma_{k}^{(V)}):\mathbb{L}_{k}\rightarrow\left\{ -1^{V}\right\} \uplus\Lambda_{k}^{(1:V)},
\]
where $1^{V}$ is the $V$-tuple of ones, $\Lambda_{k}^{(1:V)}\triangleq\Lambda_{k}^{(1)}\times\ldots\times\Lambda_{k}^{(V)}$,
and $\Lambda_{k}^{(v)}\triangleq\{0,\ldots,|Z_{k}^{(v)}|\}$. Note
that each $\gamma_{k}^{(v)}$ is required to be positive 1-1, in which
case $\gamma_{k}$ is said to be positive 1-1. Here, $\gamma_{k}(\ell)=-1^{V}$
means label $\ell$ does not exist, whereas $\gamma_{k}(\ell)\in\Lambda_{k}^{(1:V)}$
means label $\ell$ exists, in which case it is misdetected by sensor
$v$ if $\gamma_{k}^{(v)}(\ell)=0$, or generates the $\gamma_{k}^{(v)}(\ell)$-th
measurement at sensor $v$ if $\gamma_{k}^{(v)}(\ell)>0$. Since the
live label set $\mathcal{L}(\gamma_{k}^{(v)})$ can be written as
$\{\ell\in\mathbb{L}_{k}:\gamma_{k}(\ell)\in\Lambda_{k}^{(1:V)}\}$,
which is independent of $v$, we write it as $\mathcal{L}(\gamma_{k})$. 

Assuming that the sensors are independent conditional on $\boldsymbol{X}_{k}$,
the multi-sensor multi-object likelihood function is simply the product
of the likelihoods of individual sensors. Let $Z_{k}\triangleq(Z_{k}^{(1)},...,Z_{k}^{(V)})$
denote the multi-sensor observation, then the multi-sensor multi-object
likelihood function is
\begin{equation}
g_{k}(Z_{k}|\boldsymbol{X}_{k})\propto\underset{\gamma_{k}\in\Gamma_{k}}{\sum}\delta_{\mathcal{L}(\gamma_{k})}[\mathcal{L}(\boldsymbol{X}_{k})][\psi_{k,Z_{k}}^{(\gamma_{k}\circ\mathcal{L}(\cdot))}(\cdot)]^{\boldsymbol{X}_{k}},\label{eq:multi_sensor_likelihood_func}
\end{equation}
where $\Gamma_{k}$ is the space of multi-sensor association maps,
and\textcolor{blue}{
\[
{\color{black}{\color{blue}}\psi_{k,Z_{k}}^{(\alpha^{(1:V)})}(\boldsymbol{x})\triangleq\overset{V}{\underset{v=1}{\prod}}\psi_{k,Z_{k}^{(v)}}^{(v,\alpha^{(v)})}(\boldsymbol{x}).}
\]
}

\subsection{Multi-object Bayes Recursion\label{ss:Multi-Object-Bayes-Recursion}}

All information about the set of objects in the surveillance region
for the interval \{$0:k\}$ is captured by the multi-object posterior,
which can be recursively propagated in time by,
\[
\boldsymbol{\pi}_{0:k}(\boldsymbol{X}_{\negthinspace0:k})\propto g_{k}(Z_{k}|\boldsymbol{X}_{\negthinspace k})\boldsymbol{f}_{k|k-1}(\boldsymbol{X}_{\negthinspace k}|\boldsymbol{X}_{\negthinspace k-1})\boldsymbol{\pi}_{0:k-1}(\boldsymbol{X}_{\negthinspace0:k-1}).
\]
Note that the posterior is conditioned on the measurement history
$Z_{1:k}$, but we have omitted it for brevity. The dimensionality
of the posterior (hence complexity per timestep) grows with time,
and computing it at each timestep is impractical. Nevertheless, the
complexity per timestep can be capped by smoothing over short windows
and linking trajectory estimates between windows via their labels.
Multi-object filtering can be regarded as a special case with a window
length of one.

For the single-sensor special case, Particle Markov Chain Monte Carlo
\cite{Andrieu2010} has been applied to approximate the multi-object
posterior in \cite{Vu2014}. Further, an analytic solution called
the (multi-scan) GLMB filter/smoother was developed in \cite{Vo2019b},
which conceptually extends to the multi-sensor case. 

\subsection{Multi-Sensor GLMB Posterior}

Before delving into the multi-sensor GLMB posterior, it is informative
to consider the association weight and trajectory posterior of a label
$\ell$. Recall that $s(\ell)$ and $t(\ell)$ are, respectively,
the earliest and latest times on $\{0:k\}$ such that $\ell$ exists.
There are four possible cases for its trajectory: (i) new born, $s(\ell)=k$;
(ii) surviving, $t(\ell)=k>s(\ell)$; (iii) die at time $k$, $t(\ell)=k-1$;
(iv) died before time $k$, $t(\ell)<k-1$. Suppose that $\ell$ generated
the sequence $\alpha_{s(\ell):k}$ of multi-sensor measurement indices,
then its trajectory posterior at time $k$ is \allowdisplaybreaks
\begin{align}
 & \tau{}_{0:k}^{(\alpha_{s(\ell):k})}(x_{s(\ell):t(\ell)},\ell)\label{eq:paramd_multi_scan_multi_sensor_trajectory_update}\\
 & =\begin{cases}
\negthinspace\frac{\Lambda_{B,k}^{(\alpha_{k})}(x_{k},\ell)}{\bar{\Lambda}_{B,k}^{(\alpha_{k})}(\ell)}, & \negthinspace\negthinspace\negthinspace\negthinspace\negthinspace s(\ell)=k\\
\negthinspace\frac{\Lambda_{S,k|k-1}^{(\alpha_{k})}(x_{k}|x_{k-1},\ell)\tau_{0:k-1}^{(\alpha_{s(\ell):k-1})}(x_{s(\ell):k-1},\ell)}{\bar{\Lambda}_{S,k|k-1}^{(\alpha_{s(\ell):k})}(\ell)}, & \negthinspace\negthinspace\negthinspace\negthinspace\negthinspace t(\ell)=k>s(\ell)\\
\negthinspace\frac{Q_{S,k-1}(x_{k-1},\ell)\tau_{0:k-1}^{(\alpha_{s(\ell):k-1})}(x_{s(\ell):k-1},\ell)}{\bar{Q}_{S,k-1}^{(\alpha_{s(\ell):k-1})}(\ell)}, & \negthinspace\negthinspace\negthinspace\negthinspace\negthinspace t(\ell)=k-1\\
\negthinspace\tau_{0:t(\ell)}^{(\alpha_{s(\ell):t(\ell)})}(x_{s(\ell):t(\ell)},\ell), & \negthinspace\negthinspace\negthinspace\negthinspace\negthinspace t(\ell)<k-1
\end{cases},\nonumber 
\end{align}
where $\tau_{0:k}^{(\alpha_{s(\ell):k-1})}(x_{s(\ell):k-1},\ell)$
is the trajectory posterior at time $k-1$, \allowdisplaybreaks
\begin{align}
\Lambda_{B,k}^{(\alpha_{k})}(x,\ell)= & \ \psi_{k,Z_{k}}^{(\alpha_{k})}(x,\ell)P_{B,k}(\ell)f_{B,k}(x,\ell),\label{eq:paramd_multi_scan_multi_sensor_weight_update_2}\\
\bar{\Lambda}_{B,k}^{(\alpha_{k})}(\ell)= & \int\Lambda_{B,k}^{(\alpha_{k})}(x,\ell)dx,\label{eq:paramd_multi_scan_multi_sensor_weight_update_3}\\
\Lambda{}_{S,k|k-1}^{(\alpha_{k})}(x_{k}|x_{k-1},\ell)= & \ \psi_{k,Z_{k}}^{(\alpha_{k})}(x_{k},\ell)P_{S,k-1}(x_{k-1},\ell)\label{eq:paramd_multi_scan_multi_sensor_weight_update_4}\\
\times & \ f_{S,k|k-1}(x_{k}|x_{k-1},\ell),\nonumber \\
\bar{\Lambda}{}_{S,k|k-1}^{(\alpha_{s(\ell):k})}(\ell)= & \int\tau_{0:k-1}^{(\alpha_{s(\ell):k-1})}(x_{s(\ell):k-1},\ell)\label{eq:paramd_multi_scan_multi_sensor_weight_update_5}\\
\times & \ \Lambda_{S,k|k-1}^{(\alpha_{k})}(x_{k}|x_{k-1},\ell)dx_{s(\ell):k},\nonumber \\
\bar{Q}{}_{S,k-1}^{(\alpha_{s(\ell):k-1})}(\ell)= & \int\tau_{0:k-1}^{(\alpha_{s(\ell):k-1})}(x_{s(\ell):k-1},\ell)\label{eq:paramd_multi_scan_multi_sensor_weight_update_6}\\
\times & \ Q_{S,k-1}(x_{k-1},\ell)dx_{s(\ell):k-1}.\nonumber 
\end{align}
In addition, the association weight for $\ell\in\mathbb{L}_{k}$ is
defined as \allowdisplaybreaks
\begin{equation}
\eta_{k|k-1}^{(\alpha_{s(\ell):k})}(\ell)=\begin{cases}
\bar{\Lambda}_{B,k}^{(\alpha_{k})}(\ell), & \negthinspace\negthinspace\negthinspace\negthinspace s(\ell)=k\\
\bar{\Lambda}_{S,k|k-1}^{(\alpha_{s(\ell):k})}(\ell), & \negthinspace\negthinspace\negthinspace\negthinspace t(\ell)=k>s(\ell)\\
\bar{Q}_{S,k-1}^{(\alpha_{s(\ell):k-1})}(\ell), & \negthinspace\negthinspace\negthinspace\negthinspace t(\ell)=k-1\\
Q_{B,k}(\ell), & \negthinspace\negthinspace\negthinspace\negthinspace\ell\in\mathbb{B}_{k},\alpha_{k}=-1{}^{V}
\end{cases}\negthinspace\negthinspace.\label{eq:paramd_multi_scan_multi_sensor_weight_update_1}
\end{equation}

We now present the multi-object posterior as follows. Starting with
the initial prior $\boldsymbol{\pi}_{0}(\boldsymbol{X}_{0})=\delta_{0}[\mathcal{L}(\boldsymbol{X}_{0})]$
(i.e., there are no live objects at the beginning) with weight $w_{0}^{(\gamma_{0})}=1$,
the multi-sensor GLMB posterior at time $k$ can be written as \cite{Vo2019b}
\begin{align}
 & \boldsymbol{\pi}_{0:k}(\boldsymbol{\boldsymbol{X}}_{0:}{}_{k})\propto\label{eq:paramd_multi_scan_multi_sensor_posterior}\\
 & \ \ \ \ \Delta(\boldsymbol{X}_{0:k})\underset{\gamma_{0:k}}{\sum}w_{0:k}^{(\gamma_{0:k})}\delta_{\mathcal{L}(\gamma_{0:k})}[\mathcal{L}(\boldsymbol{X}_{0:k})][\tau_{0:k}^{(\gamma_{0:k}\circ\mathcal{L}(\cdot))}(\cdot)]^{\boldsymbol{X}_{0:k}},\nonumber 
\end{align}
where $\Delta(\boldsymbol{X}_{0:k})\triangleq\prod_{i=0}^{k}\Delta(\boldsymbol{X}_{i})$,
and 
\begin{align}
\negthinspace\negthinspace\negthinspace\negthinspace w_{0:k}^{(\gamma_{0:k})}= & \prod_{j=1}^{k}w_{j}^{(\gamma_{0:j})},\label{eq:paramd_full_gibbs_weight}\\
\negthinspace\negthinspace\negthinspace\negthinspace w_{j}^{(\gamma_{0:j})}= & 1_{\mathcal{F}(\mathbb{B}_{j}\uplus\mathcal{L}(\gamma_{j-1}))}(\mathcal{L}(\gamma_{j}))[\eta_{j|j-1}^{(\gamma_{0:j}(\cdot))}(\cdot)]^{\mathbb{B}_{j}\uplus\mathcal{L}(\gamma_{j-1})}.\label{eq:paramd_factor_gibbs_weight}
\end{align}

\textcolor{black}{Some of the relevant posterior statistics from the
multi-scan GLMB are \cite{Vo2019b}: }
\begin{itemize}
\item \textcolor{black}{The cardinality distribution, i.e., the distribution
of the number of trajectories is given by,
\begin{equation}
\mathrm{Pr}(|\mathcal{L}(\boldsymbol{X}_{0:k})|=n)=\sum_{\gamma_{0:k}}w_{0:k}^{(\gamma_{0:k})}\delta_{n}[|\mathcal{L}(\gamma_{0:k})|];\label{eq:ntraj_dist}
\end{equation}
$\vspace{-0.3cm}$}
\item \textcolor{black}{The cardinality distribution of births and deaths
at time $u\in\{0:k\}$ are given by,
\begin{alignat}{1}
\mathrm{Pr}(n\ \mathrm{births} & \mathrm{\ at\ time\ }u)\nonumber \\
= & \sum_{\gamma_{0:k}}w_{0:k}^{(\gamma_{0:k})}\delta_{n}[\sum_{\ell\in\mathcal{L}(\gamma_{0:k})}\delta_{u}[s(\ell)]];\label{eq:birth_dist}\\
\mathrm{Pr}(n\ \mathrm{deaths} & \mathrm{\ at\ time\ }u)\nonumber \\
= & \sum_{\gamma_{0:k}}w_{0:k}^{(\gamma_{0:k})}\delta_{n}[\sum_{\ell\in\mathcal{L}(\gamma_{0:k})}\delta_{u}[t(\ell)]];\label{eq:death_dist}
\end{alignat}
$\vspace{-0.2cm}$}
\item \textcolor{black}{The distribution of trajectory lengths is given
by,
\begin{align}
\mathrm{Pr}(\mathrm{a}\  & \mathrm{trajectory}\mathrm{\ has\ length\ }m)\nonumber \\
= & \sum_{\gamma_{0:k}}\frac{w_{0:k}^{(\gamma_{0:k})}}{|\mathcal{L}(\gamma_{0:k})|}\sum_{\ell\in\mathcal{L}(\gamma_{0:k})}\delta_{m}[t(\ell)-s(\ell)+1].\label{eq:ltraj_dist}
\end{align}
$\vspace{-0.2cm}$}
\end{itemize}
Several multi-object trajectory estimators \cite{Vo2019b} are applicable
to the GLMB posterior (\ref{eq:paramd_multi_scan_multi_sensor_posterior}).
This work uses the GLMB estimator, which first determines the most
probable cardinality $n^{*}$ by maximizing (17),
and then the highest weighted component (indexed by) $\gamma_{0:k}^{*}$
with cardinality $n^{*}$. The $n^{*}$ estimated trajectories can
be taken as the mean or mode of the trajectory posterior $\tau_{0:k}^{(\gamma_{0:k}^{*}(\ell))}(\cdot,\ell)$
for each $\ell\in\mathcal{L}(\gamma_{0:k}^{*})$.

The GLMB posterior (\ref{eq:paramd_multi_scan_multi_sensor_posterior})
is completely parameterized by the set of components $\{(w_{0:k}^{(\gamma_{0:k})},\tau_{0:k}^{(\gamma_{0:k})})\}$,
indexed by $\gamma_{0:k}$, which grow super-exponentially in number,
and hence a tractable approximation is necessary. Truncation by retaining
a prescribed number of components with highest weights minimizes the
$L_{1}$-norm approximation error \cite{Vo2019b}. 

\section{Multi-Sensor GLMB Posterior Propagation\label{sec:Comp_MSMS_GLMB_Posts}}

This section presents techniques to truncate the multi-sensor GLMB
posterior by extending the Gibbs samplers proposed in \cite{Vo2019b}.
Subsection \ref{subsec:Stationary-distribution} introduces the distributions
from which significant components of the GLMB posterior are drawn.
Subsection \ref{subsec:Sampling-from-factors} presents an efficient
algorithm to generate valid components that can be used to initialize
the full Gibbs sampler presented in Subsection \ref{subsec:The-Full-Gibbs},
which generates components according to a desired distribution. 

\subsection{Sampling Distributions\label{subsec:Stationary-distribution}}

We truncate the posterior GLMB by sampling its components from some
discrete distribution $\pi$ such that those with higher weights are
more likely to be chosen. Without loss of generality, we start with
$\mathcal{L}(\gamma_{0})=\emptyset$, and consider
\begin{align}
\pi(\gamma_{0:k})= & \ \overset{k}{\underset{j=1}{\prod}}\pi^{(j)}(\gamma_{j}|\gamma_{0:j-1}),\label{eq:stationary_dist_gamma}
\end{align}
where
\begin{align}
\pi^{(j)}(\gamma_{j}|\gamma_{0:j-1})\propto & \ 1_{\Gamma_{j}}(\gamma_{j})1_{\mathcal{F}(\mathbb{B}_{j}\uplus\mathcal{L}(\gamma_{j-1}))}(\mathcal{L}(\gamma_{j}))\label{eq:factored_dist_of_gamma}\\
\times & \ [\vartheta_{j}^{(\gamma_{0:j}(\cdot))}(\cdot)]^{\mathbb{B}_{j}\uplus\mathcal{L}(\gamma_{j-1})},\nonumber 
\end{align}
and $\vartheta_{j}^{(\gamma_{0:j}(\cdot))}(\cdot)$ is chosen to be
approximately proportional to $\eta_{j|j-1}^{(\gamma_{0:j}(\cdot))}(\cdot)$
so that (\ref{eq:stationary_dist_gamma}) is approximately proportional
to $w_{0:k}^{(\gamma_{0:k})}$ in (\ref{eq:paramd_full_gibbs_weight}).
A function $f$ is said to be approximately proportional to $g$,
i.e., $f\approxprop g$, if $f/\langle f,1\rangle\simeq g/\langle g,1\rangle.$
Note that, if $\vartheta_{j}^{(\gamma_{0:j}(\cdot))}(\cdot)$ is equal
to $\eta_{j|j-1}^{(\gamma_{0:j}(\cdot))}(\cdot)$, (\ref{eq:factored_dist_of_gamma})
is proportional to (\ref{eq:paramd_factor_gibbs_weight}), and (\ref{eq:stationary_dist_gamma})
is proportional to (\ref{eq:paramd_full_gibbs_weight}).

The term $1_{\mathcal{F}(\mathbb{B}_{j}\uplus\mathcal{L}(\gamma_{j-1}))}(\mathcal{L}(\gamma_{j}))$
in (\ref{eq:factored_dist_of_gamma}) means $\mathcal{L}(\gamma_{j})\subseteq\mathbb{B}_{j}\uplus\mathcal{L}(\gamma_{j-1})$,
i.e., $\gamma_{j}(\ell)=-1^{V}$ for all $\ell\notin\mathbb{B}_{j}\uplus\mathcal{L}(\gamma_{j-1})$.
Hence, only the values of $\gamma_{j}$ on $\mathbb{B}_{j}\uplus\mathcal{L}(\gamma_{j-1})$
need to considered. In addition, the following decomposition of $1_{\Gamma_{i}}(\gamma_{i})$
is instrumental for the proposed Gibbs samplers (the proof is given
in Supplementary Material, Subsection A).
\begin{lem}
\label{lem:multi_scan_multi_sensor_pos_1_1}Let $\bar{n}\triangleq\{1:|\mathbb{L}_{j}|\}-\{n\}$,
and $\Gamma_{j}(\bar{n})$ be the set of all $\gamma_{j}(\ell_{\bar{n}})\triangleq(\gamma_{j}(\ell_{1:n-1}),\gamma_{j}(\ell_{n+1:|\mathbb{L}_{j}|}))\in(\{-1\}^{V}\uplus\Lambda_{j}^{(1:V)})^{|\mathbb{L}_{j}|-1}$
that are positive 1-1. Then,
\begin{align}
1_{\Gamma_{j}}(\gamma_{j})= & \ 1_{\Gamma_{j}(\bar{n})}(\gamma_{j}(\ell_{\bar{n}}))\overset{V}{\underset{v=1}{\prod}}\beta_{n}^{(j,v)}(\gamma_{j}^{(v)}(\ell_{n})|\gamma_{j}^{(v)}(\ell_{\bar{n}})),\label{eq:lem_multi_sensor_positive_1_1_assignment_decomp}
\end{align}
where
\begin{align}
 & \beta_{n}^{(j,v)}(\gamma_{j}^{(v)}(\ell_{n})|\gamma_{j}^{(v)}(\ell_{\bar{n}}))=\label{eq:lem_multi_sensor_positive_1_1_assignment_decom2}\\
 & \ \ \ \ \ \ \left[1-1_{\{{\color{blue}{\color{black}1:|Z_{j}^{(v)}}|}\}\cap\{\gamma_{j}^{(v)}(\ell_{1:n-1}),\gamma_{j}^{(v)}(\ell_{n+1:|\mathbb{L}_{j}|})\}}(\gamma_{j}^{(v)}(\ell_{n}))\right].\nonumber 
\end{align}
\begin{figure*}
\begin{centering}
\subfloat{\includegraphics[scale=0.6]{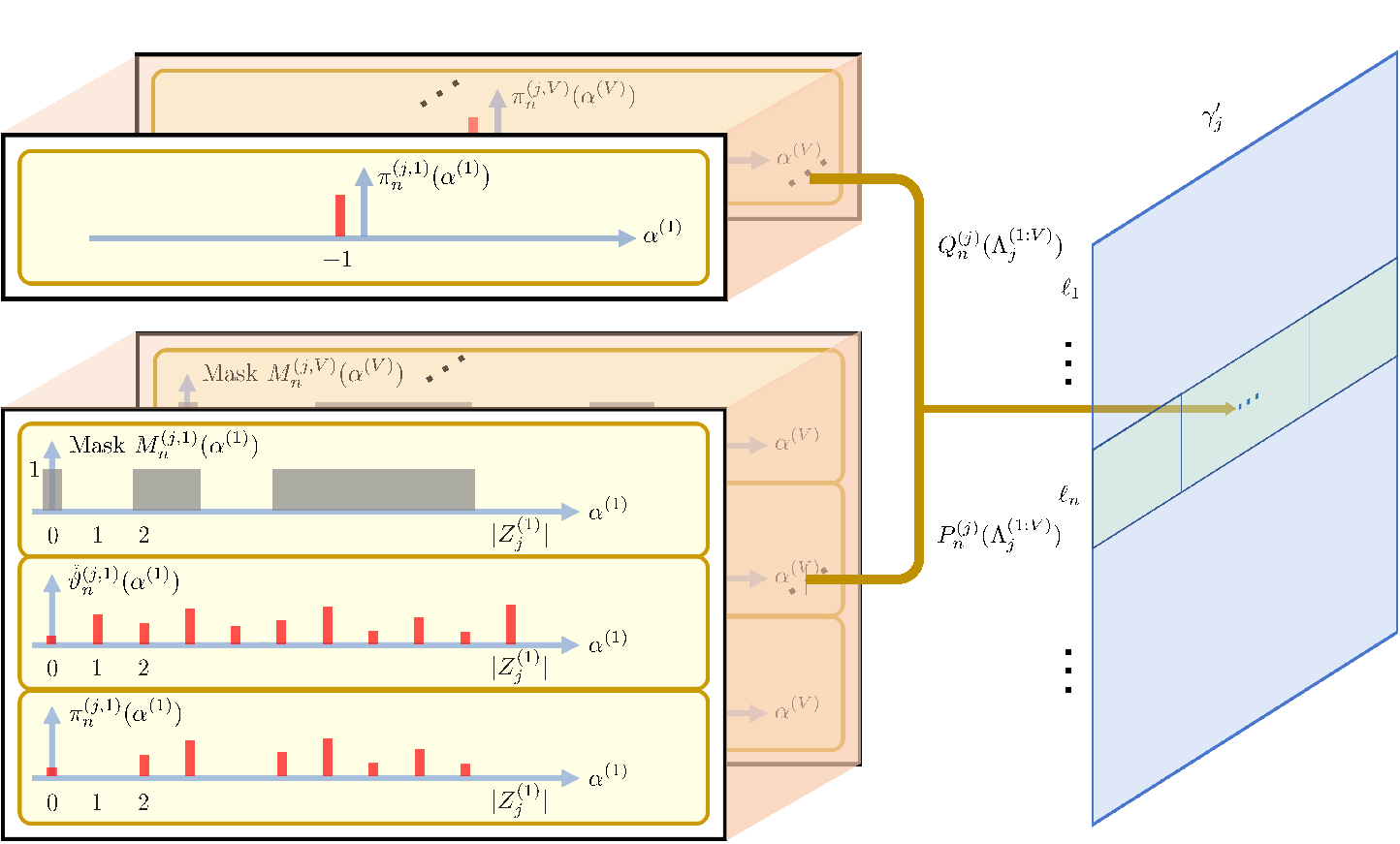}}
\par\end{centering}
\medskip{}

\begin{centering}
\begin{tabular}{|l|l|}
\hline 
$\mathring{\vartheta}_{n}^{(j,v)}(\cdot)\propto\vartheta_{j,v}(\cdot|\gamma_{0:j-1}(\ell_{n}),\ell_{n})$
given in (\ref{eq:factor_sampling_single_sensor_ass_dist}) & $\pi_{n}^{(j,1)}(\cdot)\triangleq\pi_{n}^{(j,1)}(\cdot|\gamma_{j}^{(1)}(\ell_{\bar{n}}),\gamma_{0:j-1})$
given in (\ref{eq:cor_factor_sampling_nth_component_dist_ass1})\tabularnewline
\hline 
$\pi_{n}^{(j,v)}(\cdot)\triangleq\pi_{n}^{(j,v)}(\cdot|\alpha^{(v-1)},\gamma_{j}^{(v)}(\ell_{\bar{n}}),\gamma_{0:j-1})$
given in (\ref{eq:cor_factor_sampling_nth_component_dist_assv}) & $M_{n}^{(j,v)}(\cdot)\triangleq\beta_{n}^{(j,v)}(\cdot|\gamma_{j}(\ell_{\bar{n}}))$
given in (\ref{eq:lem_multi_sensor_positive_1_1_assignment_decom2})\tabularnewline
\hline 
\end{tabular}
\par\end{centering}
\caption{\label{fig:factor_sampler}Generating $\gamma'_{j}(\ell_{n})=(\alpha^{(1)},...,\alpha^{(V)})$
from $\gamma_{j}$, $\ell_{n}\in\mathbb{B}_{j}\uplus\mathcal{L}(\gamma_{j-1})$.
Two cases: (i) bottom branch, with probability $P_{n}^{(j)}(\Lambda_{j}^{(1:V)})$,
given in (\ref{eq:cor_factor_sampling_nth_component_dist_Pn}), $\ell_{n}$
is either born (if $\ell_{n}\in\mathbb{B}_{j}$) or surviving (if
$\ell_{n}\in\mathcal{L}(\gamma_{j-1})$). For each $v\in\{1:V\}$,
$\alpha^{(v)}$ is sampled from the categorical distribution $\pi_{n}^{(j,v)}(\cdot)$
that depends on the pre-computed $\mathring{\vartheta}_{n}^{(j,v)}(\cdot)$.
To ensure that $\gamma'_{j}$ is positive 1-1, we use $M_{n}^{(j,v)}(\cdot)$
to mask out the positive measurement indices taken up by other labels
(by multiplying $\mathring{\vartheta}_{n}^{(j,v)}(\cdot)$ with $M_{n}^{(j,v)}(\cdot)$
), which results in $\pi_{n}^{(j,v)}(\cdot)$; (ii) top branch, with
probability $Q_{n}^{(j)}(\Lambda_{j}^{(1:V)})=1-P_{n}^{(j)}(\Lambda_{j}^{(1:V)})$,
$\ell_{n}$ is non-existent and the only allowable value for $\alpha^{(v)}$
is $-1$ for all $v\in\{1:V\}$.}
\end{figure*}
\end{lem}

\subsection{Sampling from the Factors \label{subsec:Sampling-from-factors}}

Sampling from (\ref{eq:stationary_dist_gamma}) can be performed by
iteratively sampling from the factors (\ref{eq:factored_dist_of_gamma}),
i.e., $\gamma_{j}\sim\pi(\cdot|\gamma_{0:j-1})$, for time $j=1:k$
(note that $\gamma_{j}$ consists of $\gamma_{j}(\ell_{n})$, $\ell_{n}\in\{\ell_{1:|\mathbb{L}_{j}|}\}\triangleq\mathbb{L}_{j}$).
We sample from $\pi(\cdot|\gamma_{0:j-1})$ using a Gibbs sampler
\cite{Casella2004}, which constructs a sequence of iterates by generating
the next iterate $\gamma'_{j}$ from $\gamma_{j}$ and the conditionals
of $\pi(\cdot|\gamma_{0:j-1})$ as follows \allowdisplaybreaks
\begin{align*}
\gamma'_{j}(\ell_{1}) & \sim\pi_{1}^{(j)}\left(\cdot|\gamma_{0:j-1},\gamma_{j}(\ell_{2:|\mathbb{L}_{j}|})\right)\\
 & \text{ \ }\vdots\\
\gamma'_{j}(\ell_{n}) & \sim\pi_{n}^{(j)}\left(\cdot|\gamma_{0:j-1},\gamma'_{j}(\ell_{1:n-1}),\gamma_{j}(\ell_{n+1:|\mathbb{L}_{j}|})\right)\\
 & \text{ \ }\vdots\\
\gamma'_{j}(\ell_{|\mathbb{L}_{j}|}) & \sim\pi_{|\mathbb{L}_{j}|}^{(j)}\left(\cdot|\gamma_{0:j-1},\gamma'_{j}(\ell_{1:|\mathbb{L}_{j}|-1})\right),
\end{align*}
where for any $\alpha=\alpha^{(1:V)}\in\{-1\}^{V}\uplus\Lambda_{j}^{(1:V)}$
\begin{align}
\pi_{n}^{(j)}(\alpha| & \gamma'_{j}(\ell_{1:n-1}),\gamma_{j}(\ell_{n+1:|\mathbb{L}_{j}|}),\gamma_{0:j-1})\nonumber \\
\propto & \ \pi^{(j)}(\gamma'_{j}(\ell_{1:n-1}),\alpha,\gamma_{j}(\ell_{n+1:|\mathbb{L}_{j}|})|\gamma_{0:j-1}),\label{eq:factor_sampling_nth_cond_dist}
\end{align}
The efficiency of Gibbs sampling hinges on the availability of conditionals
that are easy to sample from.

\textcolor{black}{We adopt the multi-sensor Gibbs sampling technique
developed in \cite{Vo2019a} to sample $\gamma'_{j}(\ell_{n})$ from
the $n$-th conditional (\ref{eq:factor_sampling_nth_cond_dist}).}
The basic steps in this technique are illustrated in Fig. \ref{fig:factor_sampler},
and summarized as pseudocode in Algorithm \ref{algo:factor_samplecoord}.
The pseudocode for generating $R$ iterates of the factor Gibbs sampler
is given in Algorithm \ref{algo:factor_gibbs}. Theoretical justifications
and analyses of Algorithms \ref{algo:factor_samplecoord} and \ref{algo:factor_gibbs}
are given in the remainder of this subsection.

Consider first the $n$-th conditional distribution by substituting
(\ref{eq:factored_dist_of_gamma}) into the right hand side of (\ref{eq:factor_sampling_nth_cond_dist}),
and noting that we are only interested in the functional dependency
on $\alpha$, \allowdisplaybreaks
\begin{align}
\pi_{n}^{(j)}(\alpha & |\gamma_{j}(\ell_{\bar{n}}),\gamma_{0:j-1})\label{eq:factor_sampling_nth_component_dist}\\
\propto & \ 1_{\Gamma_{j}}((\gamma_{j}(\ell_{1:n-1}),\alpha,\gamma_{j}(\ell_{n+1:|\mathbb{L}_{j}|})))\nonumber \\
\times & \ 1_{\mathcal{F}(\mathbb{B}_{j}\uplus\mathcal{L}(\gamma_{j-1}))}(\mathcal{L}((\gamma_{j}(\ell_{1:n-1}),\alpha,\gamma_{j}(\ell_{n+1:|\mathbb{L}_{j}|}))))\nonumber \\
\times & \ \vartheta_{j}^{((\gamma_{0:j-1}(\ell_{n}),\alpha))}(\ell_{n}),\nonumber 
\end{align}
where $\gamma_{j}(\ell_{u:v})\triangleq[\gamma_{j}(\ell_{u}),...,\gamma_{j}(\ell_{v})]$.
Note that, given a positive 1-1 multi-sensor association map, any
association map sampled from these conditionals is also positive 1-1.

In general, sampling $\alpha$ directly from (\ref{eq:factor_sampling_nth_component_dist})
is both memory and computationally expensive since $\vartheta_{j}^{((\gamma_{0:j-1}(\ell_{n}),\alpha))}(\ell_{n})$
needs to be evaluated for each of the $1+\prod_{v=1}^{V}(1+|Z_{j}^{(v)}|)$
possible values of $\alpha$. Fortunately, the computational cost
can be drastically reduced by using the strategy in \cite{Vo2019a}
via the so-called minimally-Markovian conditional distributions. 
\begin{algorithm}[th!]
\caption{Factor-SampleCoord}
\label{algo:factor_samplecoord} \SetKwInOut{Input}{Input}\SetKwInOut{Output}{Output}

{\small{}\Input{$G_{0:j-1}=(\gamma_{0:j-1},w_{0:j-1},\tau_{0:j-1}),\gamma'_{j},n$}
\Output{$\gamma'_{j}(\ell_{n})$} \vspace{0.1cm}
\hrule\vspace{0.1cm}
}{\small\par}

{\small{}\For{$v=1:V$}{ $[\vartheta_{j}^{(n,v)}(\alpha^{(v)}):=\vartheta_{j,v}(\alpha^{(v)}|\gamma_{0:j-1}(\ell_{n}),\ell_{n})]_{\alpha^{(v)}=-1}^{|Z_{j}^{(v)}|}$
via (\ref{eq:factor_sampling_single_sensor_ass_dist})\; } Compute
$P_{n}^{(j)}(\Lambda_{j}^{(1:V)})$ via (\ref{eq:cor_factor_sampling_nth_component_dist_Pn})\;
$Q_{n}^{(j)}(\Lambda_{j}^{(1:V)}):=1-P_{n}^{(j)}(\Lambda_{j}^{(1:V)})$\;
$\epsilon\sim\textsf{Categorical}([$``+", ``-"$],[P_{n}^{(j)}(\Lambda_{j}^{(1:V)}),Q_{n}^{(j)}(\Lambda_{j}^{(1:V)})])$\;
\uIf{$\epsilon=$ ``+"}{ $\gamma'_{j}(\ell_{n})\sim\textsf{SampleCoord}(P_{n}^{(j)}(\Lambda_{j}^{(1:V)}),$
$\vartheta_{j}^{(n,v)},\gamma'_{j},n)$\; } \Else{ $\gamma'_{j}(\ell_{n}):=-1*\textsf{ones}(1,V)$\;
}}{\small\par}
\end{algorithm}
\vspace{0cm}
\begin{algorithm}[th!]
\setcounter{algocf}{0} 
\global\long\def\thealgocf{\arabic{algocf}a}%
 \caption{SampleCoord}

\label{algo:sample_coord} \SetKwInOut{Input}{Input}\SetKwInOut{Output}{Output}

{\small{}\Input{$P(\Lambda_{j}^{(1:V)}),\vartheta_{j}^{(n,v)},\gamma'_{j},n$}
\Output{$[\phi_{n}^{(v)}]_{v=1}^{V}$} \vspace{0.1cm}
}{\small\par}

\hrule{\small{}\vspace{0.1cm}
}{\small\par}

{\small{}\For{$v=1:V$}{ $c^{(v)}:=[0:|Z_{j}^{(v)}|]$\; \For{$\alpha^{(v)}=0:|Z_{j}^{(v)}|$}{
$p_{j}^{(n,v)}(\alpha^{(v)}):=\vartheta_{j}^{(n,v)}(\alpha^{(v)})\beta_{n}^{(j,v)}(\alpha^{(v)}|\gamma_{j}^{(v)}(\ell_{\bar{n}}))$\;
\If{$v=1$}{ $p_{j}^{(n,v)}(\alpha^{(v)}):=p_{j}^{(n,v)}(\alpha^{(v)})P(\Lambda_{j}^{(1:V)})$\;
} } $\phi_{n}^{(v)}\sim\textsf{Categorical}(c^{(v)},[p_{j}^{(n,v)}(\alpha^{(v)})]_{v=1}^{V})$\;
}}{\small\par}
\end{algorithm}
\begin{algorithm}[h!]
\setcounter{algocf}{1} 
\global\long\def\thealgocf{\arabic{algocf}}%
 \caption{Factor-Gibbs}
\label{algo:factor_gibbs} \SetKwInOut{Input}{Input}\SetKwInOut{Output}{Output}

{\small{}\Input{$G_{0:j-1}=(\gamma_{0:j-1},w_{0:j-1},\tau_{0:j-1}),$\\$R$
(no. new samples)} \Output{$[G_{0:k}^{(r)}]_{r=1}^{R}$} \vspace{0.1cm}
}{\small\par}

\hrule

{\small{}\vspace{0.1cm}
}{\small\par}

{\small{}$P_{j}:=|\mathbb{B}_{j}\uplus\mathcal{L}(\gamma_{j-1})|$\;
$\gamma_{j}:=$ }\textsf{\small{}zeros}{\small{}$(P_{j},V)$; (or
any valid state)}\\
{\small{} \For{$r=1:R$}{ $\gamma'_{j}:=\gamma_{j}$\; \For{$n=1:P_{j}$}{
$\gamma'_{j}(\ell_{n}):=\textsf{Factor-SampleCoord}(G_{0:j-1},\gamma'_{j},n)$\;
} $\gamma_{0:j}:=[\gamma_{0:j-1},\gamma'_{j}]$\; Compute $w_{0:j},\tau_{0:j}$
from $\gamma_{0:j}$ via (15), (7)\;
$G_{0:j}^{(r)}:=(\gamma_{0:j},w_{0:j},\tau_{0:j})$\; }}{\small\par}
\end{algorithm}

\begin{defn}
The conditional (\ref{eq:factor_sampling_nth_component_dist}) is
said to be \textit{Markovian} if
\begin{align}
\vartheta_{j}^{((\gamma_{0:j-1}(\ell),\alpha))}(\ell)= & \overset{V}{\underset{v=2}{\prod}}\ \vartheta_{j,v}(\alpha^{(v)}|\alpha^{(v-1)},\gamma_{0:j-1}(\ell),\ell)\nonumber \\
\times & \ \vartheta_{j,1}(\alpha^{(1)}|\gamma_{0:j-1}(\ell),\ell),\label{eq:markov_weight_dist}
\end{align}
and \textit{minimally-Markovian} if $\vartheta_{j,v}$ can be written
in the form
\begin{align}
\negthinspace\negthinspace\negthinspace\negthinspace\negthinspace\negthinspace\negthinspace\negthinspace\negthinspace\negthinspace\negthinspace\negthinspace\vartheta_{j,v}( & \alpha^{(v)}|\alpha^{(v-1)},\gamma_{0:j-1}(\ell),\ell)=\label{eq:cor_factor_sampling_min_markov_weight_dist}\\
\vartheta_{j,v} & (\alpha^{(v)}|\gamma_{0:j-1}(\ell),\ell)1_{\{-1\}^{2}\uplus\Lambda_{j}^{(v-1:v)}}(\alpha^{(v-1)},\alpha^{(v)}).\nonumber 
\end{align}

The Markov property allows us to sample $\alpha^{(1)},\ldots,\alpha^{(V)}$
individually as shown in Fig. \ref{fig:factor_sampler}, thereby alleviating
the evaluations of $\vartheta_{j}^{((\gamma_{0:j-1}(\ell_{n}),\alpha))}(\ell_{n})$
over all possible values of $\alpha$. Better still, a complexity
of $\mathcal{O}(kV\dot{M})$, where $\dot{M}=\mathrm{max}_{j\in{1:k}}\mathrm{max}_{v\in{1:V}}\{|Z_{j}^{(v)}|\}$,
can be achieved using minimally-Markovian conditionals whose explicit
forms are given in the following Proposition (which follows from Corollary
4 of \cite{Vo2019a}).
\end{defn}
\begin{prop}
\label{prop:factor_sampling}Let $\gamma_{j}(\ell_{\bar{n}})=(\gamma_{j}(\ell_{1:n-1}),\gamma_{j}(\ell_{n+1:|\mathbb{L}_{j}|}))$
be positive 1-1, and suppose that the conditional $\pi_{n}^{(j)}(\cdot|\gamma_{j}(\ell_{\bar{n}}),\gamma_{0:j-1})$,
given by (\ref{eq:factor_sampling_nth_component_dist}) is minimally-Markovian.
Then, for $\ell_{n}\in\mathbb{L}_{j}-(\mathbb{B}_{j}\uplus\mathcal{L}(\gamma_{j-1}))$,
\begin{equation}
\pi_{n}^{(j)}(\gamma_{j}(\ell_{n})|\gamma_{j}(\ell_{\bar{n}}),\gamma_{0:j-1})=1_{\{-1\}^{V}}(\gamma_{j}(\ell_{n})){\color{blue}},\label{eq:prop_factor_sampling_nth_component_out_of_live_label_space-1}
\end{equation}
 and for $\ell_{n}\in\mathbb{B}_{j}\uplus\mathcal{L}(\gamma_{j-1})$,
\begin{align}
\pi_{n}^{(j)}(\gamma_{j} & (\ell_{n})|\gamma_{j}(\ell_{\bar{n}}),\gamma_{0:j-1})\nonumber \\
= & \left(\overset{V}{\underset{v=2}{\prod}}\pi_{n}^{(j,v)}(\gamma_{j}^{(v)}(\ell_{n})|\gamma_{j}^{(v-1)}(\ell_{n}),\gamma_{j}^{(v)}(\ell_{\bar{n}}),\gamma_{0:j-1})\right)\nonumber \\
\times & \ \pi_{n}^{(j,1)}(\gamma_{j}^{(1)}(\ell_{n})|\gamma_{j}^{(1)}(\ell_{\bar{n}}),\gamma_{0:j-1}),\label{eq:prop_factor_sampling_prop_nth_component_dist-1}
\end{align}
where
\begin{align}
\pi_{n}^{(j,1)} & (\alpha^{(1)}|\gamma_{j}^{(1)}(\ell_{\bar{n}}),\gamma_{0:j-1})\label{eq:cor_factor_sampling_nth_component_dist_ass1}\\
= & \begin{cases}
1-P_{n}^{(j)}(\Lambda_{j}^{(1:V)}), & \alpha^{(1)}=-1\\
P_{n}^{(j)}(\Lambda_{j}^{(1:V)})\beta_{n}^{(j,1)}(\alpha^{(1)}|\gamma^{(1)}(\ell_{\bar{n}}))\\
\times\frac{\vartheta_{j,1}(\alpha^{(1)}|\gamma_{0:j-1}(\ell_{n}),\ell_{n})}{\Upsilon_{n}^{(j,1)}}, & \alpha^{(1)}>-1
\end{cases}\nonumber \\
\pi_{n}^{(j,v)} & (\alpha^{(v)}|\alpha^{(v-1)},\gamma_{j}^{(v)}(\ell_{\bar{n}}),\gamma_{0:j-1})\label{eq:cor_factor_sampling_nth_component_dist_assv}\\
= & \begin{cases}
1, & \alpha^{(v-1)},\alpha^{(v)}=-1\\
\beta_{n}^{(j,v)}(\alpha^{(v)}|\gamma_{j}^{(v)}(\ell_{\bar{n}}))\\
\times\frac{\vartheta_{j,v}(\alpha^{(v)}|\gamma_{0:j-1}(\ell_{n}),\ell_{n})}{\Upsilon_{n}^{(j,v)}}, & \alpha^{(v-1)},\alpha^{(v)}>-1
\end{cases}\nonumber 
\end{align}
for $v\in\left\{ 2:V\right\} $, and
\begin{align}
\negthinspace\negthinspace\negthinspace P_{n}^{(j)}(\Lambda_{j}^{\negthinspace(1:V)}) & \triangleq\frac{\overset{V}{\underset{v=1}{\prod}}\Upsilon_{n}^{(j,v)}}{\overset{V}{\underset{v=1}{\prod}}\negthinspace\vartheta_{j,v}(-1|\gamma_{0:j-\negthinspace\negthinspace1}(\ell_{n}),\ell_{n})+\overset{V}{\underset{v=1}{\prod}}\negthinspace\Upsilon_{n}^{(j,v)}},\negthinspace\negthinspace\label{eq:cor_factor_sampling_nth_component_dist_Pn}\\
\Upsilon_{n}^{(j,v)} & \triangleq\overset{M_{j}^{(v)}}{\underset{i=0}{\sum}}\beta_{n}^{(j,v)}(i|\gamma_{j}^{(v)}(\ell_{\bar{n}}))\vartheta_{j,v}(i|\gamma_{0:j-1}(\ell_{n}),\ell_{n}).\nonumber 
\end{align}
\end{prop}
In addition to the minimally-Markovian property (for the desired computational
complexity), recall from (\ref{eq:factored_dist_of_gamma}) that 
\begin{equation}
\vartheta_{j}^{(\gamma_{0:j-1}(\ell),\alpha)}(\ell)\approxprop\eta_{j|j-1}^{(\gamma_{0:j-1}(\ell),\alpha)}(\ell),\label{eq:factor_sampling_joint_multi_sensor_ass_dist}
\end{equation}
where $\eta_{j|j-1}^{(\gamma_{0:j-1}(\ell),\alpha)}(\ell)$ is defined
in (\ref{eq:paramd_multi_scan_multi_sensor_weight_update_1}), and
can be interpreted as the (unnormalized) probability that\textcolor{red}{{}
}(conditioned on $\gamma_{0:j-1}(\ell)$ and $Z_{1:j}$)\textcolor{red}{{}
}label $\ell$ generates measurements $z_{\alpha^{(1)}}^{(1)},...,z_{\alpha^{(V)}}^{(V)}$
at time $j$ (with the understanding that $\alpha^{(v)}=0$ refers
to a miss-detection and $\alpha^{(v)}=-1$ refers to a non-existence),
abbreviated as $\text{Pr}_{j}(\ell\sim z_{\alpha^{(1)}}^{(1)},...,z_{\alpha^{(V)}}^{(V)})$.
This can be accomplished by choosing
\begin{equation}
\vartheta_{j,v}(\alpha^{(v)}|\gamma_{0:j-1}(\ell),\ell)\propto\eta_{j|j-1}^{(\gamma_{0:j-1}(\ell),(v;\alpha^{(v)}))}(\ell),\label{eq:factor_sampling_single_sensor_ass_dist}
\end{equation}
where 
\begin{align}
\negthinspace\negthinspace\negthinspace\negthinspace\eta & _{j|j-1}^{(\alpha_{s(\ell):j-1},(v;\alpha^{(v)}))}(\ell)\nonumber \\
\negthinspace\negthinspace\negthinspace\negthinspace\negthinspace\negthinspace\negthinspace & \triangleq\begin{cases}
\bar{\Lambda}_{B,j}^{(v;\alpha^{(v)})}(\ell), & \negthinspace\negthinspace\negthinspace\negthinspace s(\ell)=j\\
\bar{\Lambda}_{S,j|j-1}^{(\alpha_{s(\ell):j-1},(v;\alpha^{(v)}))}(\ell), & \negthinspace\negthinspace\negthinspace t(\ell)=j>s(\ell)\\
\bar{Q}_{S,j-1}^{(\alpha_{s(\ell):j-1})}(\ell), & \negthinspace\negthinspace\negthinspace\negthinspace t(\ell)=j-1\\
Q_{B,j}(\ell), & \negthinspace\negthinspace\negthinspace\negthinspace\ell\in\mathbb{B}_{j},\alpha^{(v)}=-1
\end{cases}\negthinspace,\label{eq:paramd_multi_scan_multi_sensor_weight_update_sensor_v}
\end{align}
\begin{align}
\bar{\Lambda}_{B,j}^{(v;\alpha^{(v)})}(\ell)=\negthinspace & \int\negthinspace\psi_{j,Z_{j}^{(v)}}^{(v,\alpha^{(v)})}(x,\ell)P_{B,j}(\ell)f_{B,j}(x,\ell)dx,\label{eq:paramd_multi_scan_multi_sensor_weight_update_6_sensor_v}
\end{align}
\begin{align}
\negthinspace\bar{\Lambda} & _{S,j|j-1}^{(\alpha_{s(\ell):j-1},(v;\alpha^{(v)}))\negthinspace}(\ell)=\negthinspace\int\negthinspace\tau_{0:j-1}^{(\alpha_{s(\ell):j-1})\negthinspace}(x_{s(\ell):j-1},\ell)\,\times\label{eq:paramd_multi_scan_multi_sensor_weight_update_5_sensor_v}\\
 & P_{S,j-1}(x_{j-1},\ell)f_{S,j|j-1}(x_{j}|x_{j-1},\ell)\psi_{j,Z_{j}^{(v)}}^{(v,\alpha^{(v)})}(x_{j},\ell)dx_{s(\ell):j},\negthinspace\negthinspace\negthinspace\negthinspace\nonumber 
\end{align}
$s(\ell)$ and $t(\ell)$ in (\ref{eq:paramd_multi_scan_multi_sensor_weight_update_sensor_v})
are, respectively, the earliest and latest times on $\{0:j\}$ such
that $\ell$ exists, and ${\color{black}\tau_{0:j-1}^{(\alpha_{s(\ell):j-1})}(x_{s(\ell):j-1},\ell)}$
is the trajectory posterior of ${\color{black}\ell}$ at time $j-1$.
Eq. (\ref{eq:paramd_multi_scan_multi_sensor_weight_update_sensor_v})
can be interpreted as the (unnormalized) probability that (conditioned
on $\gamma_{0:j-1}(\ell)$ and $Z_{1:j}$) label $\ell$ generates
measurement $z_{\alpha^{(v)}}^{(v)}$ at time $j$, abbreviated as
$\text{Pr}_{j}(\ell\sim z_{\alpha^{(v)}}^{(v)})$.

The rationale for choosing (\ref{eq:factor_sampling_single_sensor_ass_dist})
can be seen by substituting (\ref{eq:factor_sampling_single_sensor_ass_dist})
into (\ref{eq:cor_factor_sampling_min_markov_weight_dist}) and (\ref{eq:markov_weight_dist}),
which gives
\begin{equation}
\vartheta_{j}^{((\gamma_{0:j-1}(\ell),\alpha))}(\ell)\propto\overset{V}{\underset{v=1}{\prod}}\eta_{j|j-1}^{(\gamma_{0:j-1}(\ell),(v;\alpha^{(v)}))}(\ell),\label{eq:factor_sampling_sub_opt_joint_weight_dist}
\end{equation}
i.e., $\vartheta_{j}^{((\gamma_{0:j-1}(\ell),\alpha))}(\ell)$ is
the product of the probabilities $\text{Pr}_{j}(\ell\sim z_{\alpha^{(v)}}^{(v)})$,
$v=1:V$. Consequently, the approximation in (\ref{eq:factor_sampling_joint_multi_sensor_ass_dist})
boils down to approximating $\text{Pr}_{j}(\ell\sim z_{\alpha^{(1)}}^{(1)},...,z_{\alpha^{(V)}}^{(V)})$
by $\text{Pr}_{j}(\ell\sim z_{\alpha^{(1)}}^{(1)})\times...\times\text{Pr}_{j}(\ell_{n}\sim z_{\alpha^{(V)}}^{(V)}).$
This is reasonable, because the event space is $\alpha\in\{-1\}^{V}\uplus\Lambda_{j}^{(1:V)}$,
and conditioned on $Z_{1:j}$ and $\gamma_{0:j-1}$, the probability
that $\ell$ generating a measurement from one sensor is almost independent
from generating a measurement from another sensor. Furthermore, it
follows from \cite{Vo2019a} that, the support of (\ref{eq:factored_dist_of_gamma})
with the minimally-Markovian property contains the support of (\ref{eq:paramd_factor_gibbs_weight})\textcolor{black}{.
Algorithm \ref{algo:factor_gibbs} uses the minimally-Markovian strategy
above.}

Note that although this approach produces valid multi-sensor association
history $\gamma_{0:k}$, in order to produce samples distributed according
to (\ref{eq:stationary_dist_gamma}), the Gibbs sampler for each factor
$\pi_{n}^{(j)}$ needs to be iterated for a sufficiently long time
(i.e., executing Algorithm \ref{algo:factor_gibbs} with a large $R$)
before proceeding to the next timestep. While this is not efficient,
sampling from the factors can provide a cheap way to generate valid
$\gamma_{0:k}$ to initialize the full Gibbs sampler (for faster convergence)
presented in the next subsection. 

For completeness, the pseudocode to generate a set of valid $\gamma_{0:k}$
is given in\textcolor{black}{{} Algorithm \ref{algo:factor_samplejoint},}
where the factor sampler (Algorithm \ref{algo:factor_gibbs}) is used
to sample $\gamma_{j}$ for times $j=1:k$. Let $M_{j}\triangleq\mathrm{max}_{v\in\{1:V\}}\{|Z_{j}^{(v)}|\}$,
$P_{j}\triangleq|\mathbb{B}_{j}\uplus\mathcal{L}(\gamma_{j-1})|$,
and $R$ be the number of new samples generated at time $j$. Then,
the complexity of this algorithm is $\mathcal{O}(R\sum_{j=1}^{k}P_{j}^{2}VM_{j})$
(although it can be implemented in $\mathcal{O}(R\sum_{j=1}^{k}P_{j}(P_{j}+VM_{j}))$
in practice by pre-calculating $\beta_{n}^{(k,v)}$ values). Denoting
$P\triangleq\mathrm{max}_{j\in\{1:k\}}{\color{blue}{\color{black}\{P_{j}\}}}$,
and $\dot{M}\triangleq\mathrm{max}_{j\in\{1:k\}}{\color{blue}{\color{black}\{M_{j}\}}}$,
indicatively, the complexity of the algorithm is $\mathcal{O}(kRP^{2}V\dot{M})$.
The Gibbs sampler described in\textcolor{blue}{{} $\textsf{{\color{black}{\color{blue}}Factor-Gibbs}}$
}\textcolor{black}{(pseudocode given in Algorithm \ref{algo:factor_gibbs})
}converges to the stationary distribution (\ref{eq:factored_dist_of_gamma})
at an exponential rate \cite[Proposition 2]{Vo2019a}.
\begin{algorithm}[th!]
\caption{Factor-SampleJoint}
\label{algo:factor_samplejoint} \SetKwInOut{Input}{Input}\SetKwInOut{Output}{Output}

{\small{}\Input{$Q$ (max samples), $R$} \Output{$[G_{0:k}^{(q)}]_{q=1}^{Q}$}
\vspace{0.1cm}
\hrule\vspace{0.1cm}
}{\small\par}

{\small{}Initialize $G_{0}^{(1)}=(\gamma_{0},w_{0},\tau_{0})$\;
\For{$j=1:k$}{ $Q_{0}=\textsf{min}(Q,\textsf{size}(G_{0,j-1}))$\;\For{$q=1:Q_{0}$}{
$[G_{0:j}^{(q,r)}]_{r=1}^{R}:=\textsf{Factor-Gibbs}(G_{0:j-1}^{(q)},R)$\;
} $[G_{0:j}^{(q)}]_{q=1}^{Q}:=Q$ best of $[G_{0:j}^{(q,r)}]_{(q,r)=(1,1)}^{(Q_{0},R)}$\;
Normalize weights $[w_{0:j}^{(q)}]_{q=1}^{Q}$\; }}{\small\par}
\end{algorithm}

\subsection{The Full Gibbs Sampler \label{subsec:The-Full-Gibbs}}

Instead of sampling from each factor, a full Gibbs sampler for (\ref{eq:stationary_dist_gamma})
constructs a sequence of iterates, where the next iterate $\gamma'_{0:k}$
is generated from the current $\gamma_{0:k}$ by sampling $\gamma'_{j}(\ell_{n})$
from the conditional
\begin{align}
\pi_{j,n}(\alpha| & \overbrace{\gamma'_{0:j-1}}^{\mathrm{past}},\overbrace{\gamma'_{j}(\ell_{1:n-1})}^{\mathrm{current\ (processed)}},\overbrace{\gamma_{j}(\ell_{n+1:|\mathbb{L}_{j}|})}^{\mathrm{current\ (unprocessed)}},\overbrace{\gamma_{j+1:k}}^{\mathrm{future}})\nonumber \\
\propto & \ \pi(\gamma'_{0:j-1},\gamma'_{j}(\ell_{1:n-1}),\alpha,\gamma_{j}(\ell_{n+1:|\mathbb{L}_{j}|}),\gamma_{j+1:k}).\label{eq:full_gibbs_sampling_nth_cond_dist}
\end{align}
for each $j\in\{1:k\}$, $\ell_{n}\in\{\ell_{1:|\mathbb{L}_{j}|}\}$. 

\begin{figure*}
\begin{centering}
\subfloat{\includegraphics[scale=0.6]{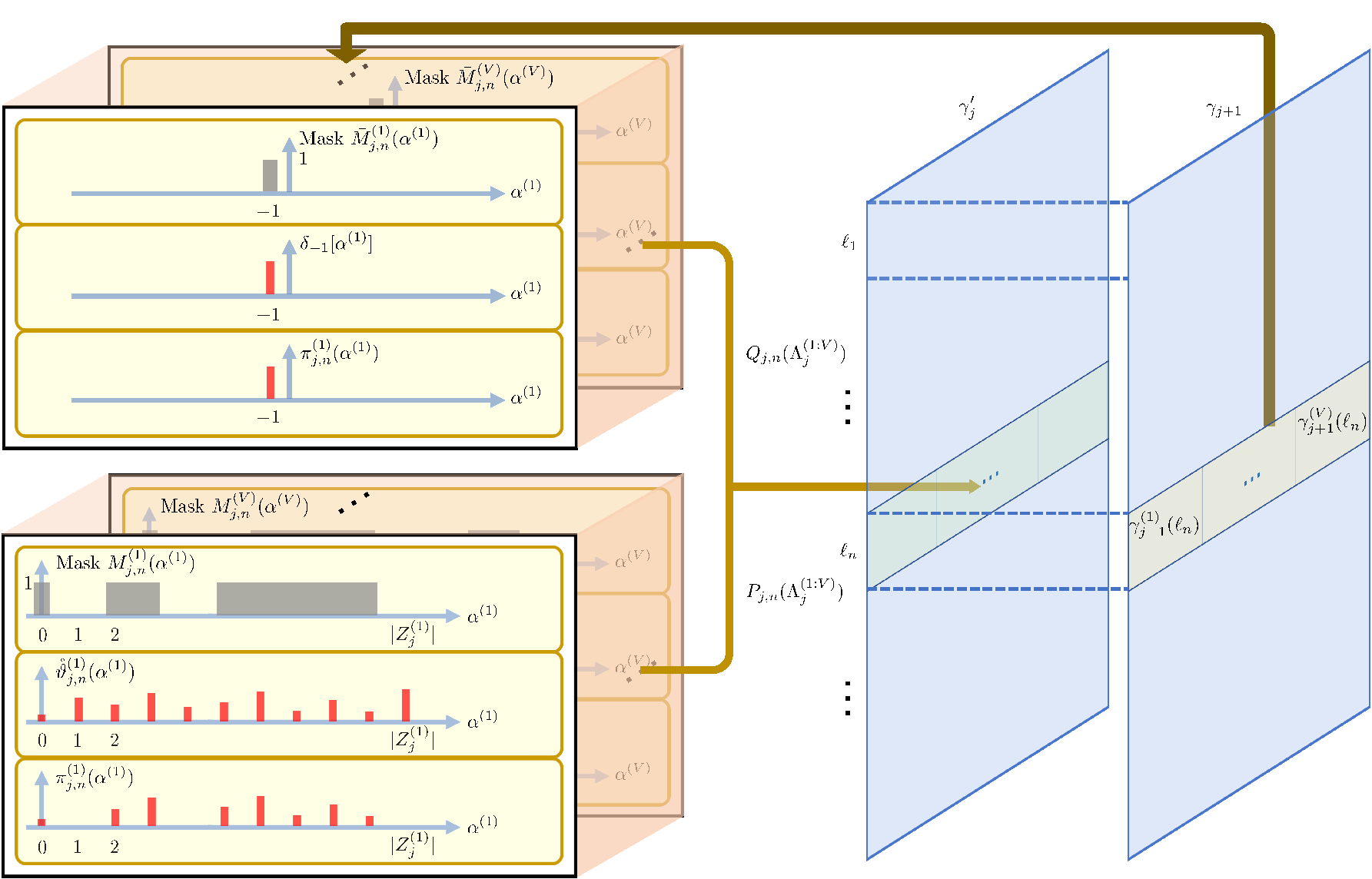}}
\par\end{centering}
\medskip{}

\begin{centering}
\begin{tabular}{|l|l|l|}
\hline 
\negthinspace{}$\mathring{\vartheta}_{j,n}^{(v)}(\cdot)\propto\vartheta_{j,v}(\cdot|\gamma_{\bar{j}}(\ell),\ell_{n})$
given in (\ref{eq:full_gibbs_sampling_sub_opt_indiv_ass_weight_dist}) & $\pi_{j,n}^{(1)}(\cdot)\triangleq\pi_{j,n}^{(1)}(\cdot|\gamma_{j}(\ell_{\bar{n}}),\gamma_{\bar{j}})$
given in (\ref{eq:cor_full_gibbs_sampling_nth_component_dist_ass1}) & $\bar{M}_{j,n}^{(v)}(\cdot)\triangleq\delta_{\gamma_{j+1}^{(v)}(\ell_{n})}[\cdot]$\negthinspace{}\tabularnewline
\hline 
\negthinspace{}$\pi_{j,n}^{(v)}(\cdot)\triangleq\pi_{j,n}^{(v)}(\cdot|\alpha^{(v-1)},\gamma_{j}(\ell_{\bar{n}}),\gamma_{\bar{j}})$
given in (\ref{eq:cor_full_gibbs_sampling_nth_component_dist_assv}) & $M_{j,n}^{(v)}(\cdot)\triangleq\beta_{n}^{(j,v)}(\cdot|\gamma_{j}(\ell_{\bar{n}}))$
given in (\ref{eq:lem_multi_sensor_positive_1_1_assignment_decom2}) & \tabularnewline
\hline 
\end{tabular}
\par\end{centering}
\centering{}\caption{\label{fig:full_gibbs_sampler}Generating $\gamma'_{j}(\ell_{n})=(\alpha^{(1)},...,\alpha^{(V)})$
from $\gamma_{j}$, $j\in\{1:k\}$, $\ell_{n}\in\mathbb{B}_{j}\uplus\mathcal{L}(\gamma_{j-1})$.
Two cases: (i) bottom branch, with probability $P_{j,n}(\Lambda_{j}^{(1:V)})$,
given in (\ref{eq:cor_full_gibbs_sampling_nth_component_dist_Pn}),
$\ell_{n}$ is either born (if $\ell_{n}\in\mathbb{B}_{j}$) or surviving
(if $\ell_{n}\in\mathcal{L}(\gamma_{j-1})$). For each $v\in\{1:V\}$,
$\alpha^{(v)}$ is sampled from the categorical distribution $\pi_{j,n}^{(v)}(\cdot)$
that depends on the pre-computed $\mathring{\vartheta}_{j,n}^{(v)}(\cdot)$.
To ensure that $\gamma'_{j}$ is positive 1-1 (and $\gamma'_{0:k}\in\Gamma_{0:k}$),
we mask out the positive indices of sensor $v$ that have already
been allocated to other labels (by multiplying $\mathring{\vartheta}_{j,n}^{(v)}(\cdot)$
with the mask $M_{j,n}^{(v)}(\cdot)$) to produce $\pi_{j,n}^{(v)}(\cdot)$;
(ii) top branch, with probability $Q_{j,n}(\Lambda_{j}^{(1:V)})=1-P_{j,n}(\Lambda_{j}^{(1:V)})$,
$\ell_{n}$ is non-existent, and the only allowable value for $\alpha^{(v)}$
is $-1$ for all $v\in\{1:V\}$. In addition, to ensure $\gamma'_{0:k}\in\Gamma_{0:k}$,
a non-existent $\ell_{n}$ at time $j$ must remain non-existent thereafter.
Therefore, $\pi_{j,n}^{(v)}(\cdot)$ must be $0$ if $\gamma_{j+1}^{(v)}(\ell_{n})$
is non-negative. This is accomplished by multiplying $\delta_{-1}[\cdot]$
with the mask $\bar{M}_{j,n}^{(v)}(\cdot)$. }
\end{figure*}

The proposed algorithm for sampling $\gamma'_{j}(\ell_{n})$ from
the conditional $\pi_{j,n}$ is illustrated in Fig. \ref{fig:full_gibbs_sampler},
and summarized as pseudocode in Algorithm \ref{algo:full_samplecoord}.
The pseudocode for generating $T$ iterates of the full Gibbs sampler
is given in Algorithm \ref{algo:full_gibbs}. Theoretical justifications
and analyses of Algorithm \ref{algo:full_samplecoord} and \ref{algo:full_gibbs}
are given in the remainder of this subsection. Mathematical proofs
are given in the Supplementary Material. 

\begin{lem}
\label{lem:full_gibbs_nth_cond_dev}Let $\gamma_{\bar{j}}\triangleq(\gamma_{0:j-1},\gamma_{j+1:k})$,
\textcolor{black}{and} $a\lor b$ denotes $\text{min}\{a,b\}$. Then,
the conditional (\ref{eq:full_gibbs_sampling_nth_cond_dist}) can
be expressed as \allowdisplaybreaks
\begin{align}
\pi_{j} & _{,n}(\alpha|\gamma_{j}(\ell_{\bar{n}}),\gamma_{\bar{j}})\label{eq:full_gibbs_sampling_nth_cond_dist3}\\
\propto & \ 1_{\Gamma_{j}}((\gamma_{j}(\ell_{1:n-1}),\alpha,\gamma_{j}(\ell_{n+1:|\mathbb{L}_{j}|})))\nonumber \\
\times & \ 1_{\mathcal{F}(\mathbb{B}_{j}\uplus\mathcal{L}(\gamma_{j-1}))}(\mathcal{L}((\gamma_{j}(\ell_{1:n-1}),\alpha,\gamma_{j}(\ell_{n+1:|\mathbb{L}_{j}|}))))\nonumber \\
\times & \ 1_{\mathcal{F}(\mathbb{B}_{j+1}\uplus\mathcal{L}((\gamma_{j}(\ell_{1:n-1}),\alpha,\gamma_{j}(\ell_{n+1:|\mathbb{L}_{j}|}))))}(\mathcal{L}(\gamma_{j+1}))\nonumber \\
\times & \ \vartheta_{j}(\alpha|\gamma_{\bar{j}}(\ell_{n}),\ell_{n}).\nonumber 
\end{align}
where
\begin{equation}
\vartheta_{j}(\alpha|\gamma_{\bar{j}}(\ell_{n}),\ell_{n})\triangleq\negthinspace\negthinspace\overset{k\lor(t(\ell_{n})+1)}{\underset{i=j}{\prod}}\negthinspace\negthinspace\vartheta_{i}^{(\gamma_{0:j-1}(\ell_{n}),\alpha,\gamma_{j+1:i}(\ell_{n}))}(\ell_{n}).\negthinspace\negthinspace\negthinspace\label{eq:full_gibbs_sampling_joint_weight_dist}
\end{equation}
\end{lem}
Similar to sampling from the factors, in general, sampling $\alpha$
from the conditional (\ref{eq:full_gibbs_sampling_nth_cond_dist3})
is both memory and computationally expensive since $\vartheta_{j}(\alpha|\gamma_{\bar{j}}(\ell_{n}),\ell_{n})$
needs to be evaluated for each of the $1+\prod_{v=1}^{V}(1+|Z_{j}^{(v)}|)$
) possible values of $\alpha$. Further, since each $\vartheta_{j}(\alpha|\gamma_{\bar{j}}(\ell_{n}),\ell_{n})$
is a product of terms that need to be evaluated for $i=j:k\lor(t(\ell_{n})+1)$,
it is more computationally expensive than the factor sampler. Again,
the computational cost can be drastically reduced using minimally-Markovian
conditional distributions.
\begin{defn}
The conditional (\ref{eq:full_gibbs_sampling_nth_cond_dist3}) is
said to be \textit{Markovian} if
\begin{align}
\vartheta_{j}(\alpha|\gamma_{\bar{j}}(\ell),\ell)= & \overset{V}{\underset{v=2}{\prod}}\ \vartheta_{j,v}(\alpha^{(v)}|\alpha^{(v-1)},\gamma_{\bar{j}}(\ell),\ell)\nonumber \\
\times & \ \vartheta_{j,1}(\alpha^{(1)}|\gamma_{\bar{j}}(\ell),\ell),\label{eq:markov_weight_dist_full_gibbs}
\end{align}
and \textit{minimally-Markovian} if $\vartheta_{j,v}$ can be written
in the form
\begin{align}
\negthinspace\negthinspace\negthinspace\negthinspace\negthinspace\negthinspace\negthinspace\negthinspace\negthinspace\negthinspace\negthinspace\negthinspace\vartheta_{j,v}( & \alpha^{(v)}|\alpha^{(v-1)},\gamma_{\bar{j}}(\ell),\ell)=\label{eq:full_gibbs_sampling_min_markov_weight_dist-1}\\
\vartheta_{j,v} & (\alpha^{(v)}|\gamma_{\bar{j}}(\ell),\ell)1_{\{-1\}^{2}\uplus\Lambda_{j}^{(v-1:v)}}(\alpha^{(v-1)},\alpha^{(v)}).\nonumber 
\end{align}
\end{defn}
The Markovian conditional allows us to sample individual $\alpha^{(1)},\alpha^{(2)},...,\alpha^{(V)}$,
alleviating the evaluation of $\vartheta_{j}(\alpha|\gamma_{\bar{j}}(\ell_{n}),\ell_{n})$
over all possible values of $\alpha$, but still incurs a total complexity
of $\mathcal{O}(kV\dot{M}^{2}|t(\ell_{n})-s(\ell_{n})|)$ for computing
normalization constants. Further, a linear complexity of $\mathcal{O}(kV\dot{M}|t(\ell_{n})-s(\ell_{n})|)$
can be achieved using the minimally-Markovian conditional given in
the following Proposition, which extends Corollary 4 of \cite{Vo2019a}
to multi-scan. 
\begin{algorithm}[h!]
\caption{Full-SampleCoord}
\label{algo:full_samplecoord} \SetKwInOut{Input}{Input}\SetKwInOut{Output}{Output}

{\small{}\Input{$\gamma_{0:k},\gamma'_{j},j,n$} \Output{$\gamma'_{j}(\ell_{n})$}
\vspace{0.1cm}
}{\small\par}

\hrule

{\small{}\vspace{0.1cm}
}{\small\par}

{\small{}\For{$v=1:V$}{ $[\vartheta_{j}^{(n,v)}(\alpha^{(v)}):=\vartheta_{j,v}(\alpha^{(v)}|\gamma_{\bar{j}}(\ell_{n}),\ell_{n})]_{\alpha^{(v)}=-1}^{|Z_{j}^{(v)}|}$
via (\ref{eq:full_gibbs_sampling_sub_opt_indiv_ass_weight_dist})\;
} Compute $P_{j,n}(\Lambda_{j}^{(1:V)})$ via (\ref{eq:cor_full_gibbs_sampling_nth_component_dist_Pn})\;
$Q_{j,n}(\Lambda_{j}^{(1:V)}):=1-P_{j,n}(\Lambda_{j}^{(1:V)})$\;
$\epsilon\sim\textsf{Categorical}([$``+", ``-"$],[P_{j,n}(\Lambda_{j}^{(1:V)}),Q_{j,n}(\Lambda_{j}^{(1:V)})])$\;
\uIf{$\epsilon=$ ``+"}{ $\gamma'_{j}(\ell_{n})\sim\textsf{SampleCoord}(P_{j,n}(\Lambda_{j}^{(1:V)}),$
$\vartheta_{j}^{(n,v)},\gamma'_{j},n)$\; } \Else{ $\gamma'_{j}(\ell_{n}):=-1*\textsf{ones}(1,V)$\;
}}{\small\par}
\end{algorithm}
\begin{algorithm}[h!]
\caption{Full-Gibbs}
\label{algo:full_gibbs} \SetKwInOut{Input}{Input}\SetKwInOut{Output}{Output}

{\small{}\Input{$G_{0:k}=(\gamma_{0:k},w_{0:k},\tau_{0:k}),T$ (no.
samples)} \Output{$[G_{0:k}^{(t)}]_{t=1}^{T}$} \vspace{0.1cm}
}{\small\par}

\hrule

{\small{}\vspace{0.1cm}
}{\small\par}

{\small{}\For{$t=1:T$}{ \For{$j=1:k$}{ $P_{j}:=|\mathbb{B}_{j}\uplus\mathcal{L}(\gamma_{j-1})|$;
\phantom{+} $\gamma'_{j}:=\gamma_{j}$\; \For{$n=1:P_{j}$}{
$\gamma'_{j}(\ell_{n}):=\textsf{Full-SampleCoord}(\gamma_{0:k},\gamma'_{j},j,n)$\;
} $\gamma_{0:j}:=[\gamma_{0:j-1},\gamma'_{j}]$\; Compute $w_{0:j},\tau_{0:j}$
from $\gamma_{0:j}$ via (15), (7)\;
} $G_{0:k}^{(t)}:=(\gamma_{0:k},w_{0:k},\tau_{0:k})$\; }}{\small\par}
\end{algorithm}

\begin{prop}
\label{prop:full_gibbs_sampler}Consider $\gamma_{j}$ of a valid
association history $\gamma_{0:k}$, $j\in\{1:k\}$, and define \textcolor{blue}{
\[
{\color{black}{\color{blue}}M_{j}(\alpha;\beta)\triangleq\begin{cases}
\delta_{\beta}[\alpha], & \alpha\in\{-1\}^{V}\\
1, & \alpha\in\Lambda_{j}^{(1:V)}
\end{cases}}.
\]
}Suppose that the conditional $\pi_{j,n}(\alpha|\gamma_{j}(\ell_{\bar{n}}),\gamma_{\bar{j}})$
given by (\ref{eq:full_gibbs_sampling_nth_cond_dist3}) is minimally-Markovian.
Then, for $\ell_{n}\in\mathbb{L}_{j}-(\mathbb{B}_{j}\uplus\mathcal{L}(\gamma_{j-1})),$
\begin{align}
\pi_{j,n}(\gamma_{j}(\ell_{n} & )|\gamma_{j}(\ell_{\bar{n}}),\gamma_{\bar{j}})\label{eq:prop_full_gibbs_sampling_prop_dead_label_space_dist-1}\\
= & \ 1_{\{-1\}^{V}}(\gamma_{j}(\ell_{n}))\delta_{\gamma_{\mathrm{min}\{j+1,k\}}(\ell_{n})}[\gamma_{j}(\ell_{n})],\nonumber 
\end{align}
and for $\ell_{n}\in\mathbb{B}_{j}\uplus\mathcal{L}(\gamma_{j-1})$
\allowdisplaybreaks
\begin{align}
\pi_{j,n}( & \gamma_{j}(\ell_{n})|\gamma_{j}(\ell_{\bar{n}}),\gamma_{\bar{j}})\label{eq:prop_full_gibbs_sampling_prop_nth_component_dist-1}\\
= & \left(\overset{V}{\underset{v=2}{\prod}}\pi_{j,n}^{(v)}(\gamma_{j}^{(v)}(\ell_{n})|\gamma_{j}^{(v-1)}(\ell_{n}),\gamma_{j}(\ell_{\bar{n}}),\gamma_{\bar{j}})\right)\nonumber \\
\times & \ \pi_{j,n}^{(1)}(\gamma_{j}^{(1)}(\ell_{n})|\gamma_{j}(\ell_{\bar{n}}),\gamma_{\bar{j}}){\color{blue}{\color{black}M_{j}(\gamma_{j}(\ell_{n});\gamma_{\mathrm{min}\{k,j+1\}}(\ell_{n}))}},\nonumber 
\end{align}
where 
\begin{align}
\pi_{j,n}^{(1)} & (\alpha^{(1)}|\gamma_{j}(\ell_{\bar{n}}),\gamma_{\bar{j}})\label{eq:cor_full_gibbs_sampling_nth_component_dist_ass1}\\
= & \begin{cases}
1-P_{j,n}(\Lambda_{j}^{(1:V)}), & \alpha^{(1)}=-1\\
P_{j,n}(\Lambda_{j}^{(1:V)})\beta_{n}^{(j,1)}(\alpha^{(1)}|\gamma_{j}^{(1)}(\ell_{\bar{n}}))\\
\times\frac{\vartheta_{j,1}(\alpha^{(1)}|\gamma_{\bar{j}}(\ell_{n}),\ell_{n})}{\Upsilon_{j,n}^{(1)}}, & \alpha^{(1)}>-1
\end{cases}\nonumber \\
\pi_{j,n}^{(v)} & (\alpha^{(v)}|\alpha^{(v-1)},\gamma_{j}(\ell_{\bar{n}}),\gamma_{\bar{j}})\label{eq:cor_full_gibbs_sampling_nth_component_dist_assv}\\
= & \begin{cases}
1, & \alpha^{(v-1)},\alpha^{(v)}=-1\\
\beta_{n}^{(j,v)}(\alpha^{(v)}|\gamma_{j}^{(v)}(\ell_{\bar{n}}))\\
\times\frac{\vartheta_{j,v}(\alpha^{(v)}|\gamma_{\bar{j}}(\ell_{n}),\ell_{n})}{\Upsilon_{j,n}^{(v)}}, & \alpha^{(v-1)},\alpha^{(v)}>-1
\end{cases}\nonumber 
\end{align}
for $v\in\left\{ 2:V\right\} $, and
\begin{align}
P_{j,n}(\Lambda_{j}^{(1:V)})\triangleq & \frac{\overset{V}{\underset{v=1}{\prod}}\Upsilon_{j,n}^{(v)}}{\overset{V}{\underset{v=1}{\prod}}\vartheta_{j,v}(-1|\gamma_{\bar{j}}(\ell_{n}),\ell_{n})+\overset{V}{\underset{v=1}{\prod}}\Upsilon_{j,n}^{(v)}},\label{eq:cor_full_gibbs_sampling_nth_component_dist_Pn}\\
\Upsilon_{j,n}^{(v)}\triangleq & \negthinspace\negthinspace\overset{|Z_{j}^{(v)}|}{\underset{\alpha^{(v)}=0}{\sum}\negthinspace\negthinspace}\beta_{n}^{(j,v)}(\alpha^{(v)}|\gamma_{j}^{(v)}(\ell_{\bar{n}}))\vartheta_{j,v}(\alpha^{(v)}|\gamma_{\bar{j}}(\ell_{n}),\ell_{n}).\nonumber 
\end{align}
 
\end{prop}
In addition to the minimally-Markovian property, we require $\vartheta_{j}^{(\gamma_{0:j}(\ell))}(\ell)$
to be approximately proportional to $\eta_{j|j-1}^{(\gamma_{0:j}(\ell))}(\ell)$
(see (\ref{eq:factored_dist_of_gamma})). Since $\vartheta_{j}(\alpha|\gamma_{\bar{j}}(\ell_{n}),\ell)$
in (\ref{eq:full_gibbs_sampling_nth_cond_dist3}) is a product of
$\vartheta_{i}^{(\gamma_{0:j-1}(\ell),\alpha,\gamma_{j+1:i}(\ell))}(\ell)$'s,
$i=j:k\lor(t(\ell)+1)$, this translates to
\begin{equation}
\vartheta_{i}^{(\gamma_{0:j-1}(\ell),\alpha,\gamma_{j+1:i}(\ell))}(\ell)\approxprop\eta_{i|i-1}^{(\gamma_{0:j-1}(\ell),\alpha,\gamma_{j+1:i}(\ell))}(\ell)\label{eq:full_gibbs_sampling_joint_multi_sensor_ass_dist}
\end{equation}
for each $i$, where $\eta_{i|i-1}^{(\gamma_{0:j-1}(\ell),\alpha,\gamma_{j+1:i}(\ell))}(\ell)$
can be interpreted as the (unnormalized) probability that (conditioned
on $\gamma_{\bar{j}}(\ell)$ and $Z_{1:k}$) label $\ell$ generates
measurements $z_{\alpha^{(1)}}^{(1)},...,z_{\alpha^{(V)}}^{(V)}$
at time $j$, abbreviated as $\text{Pr}_{j}(\ell_{n}\sim z_{\alpha^{(1)}}^{(1)},...,z_{\alpha^{(V)}}^{(V)})$.
To fulfill (\ref{eq:full_gibbs_sampling_joint_multi_sensor_ass_dist})
and the minimally-Markovian property, we choose
\begin{equation}
\negthinspace\vartheta_{j,v}(\alpha^{(v)}|\gamma_{\bar{j}}(\ell),\ell)\propto\negthinspace\negthinspace\negthinspace\negthinspace\overset{k\lor(t(\ell)+1)}{\underset{i=j}{\prod}}\negthinspace\negthinspace\negthinspace\eta_{i|i-1}^{(\gamma_{0:j-1}(\ell),(v;\alpha^{(v)}),\gamma_{j+1:i}(\ell))}(\ell),\negthinspace\negthinspace\label{eq:full_gibbs_sampling_sub_opt_indiv_ass_weight_dist}
\end{equation}
where
\begin{align}
\negthinspace\negthinspace\negthinspace\negthinspace\eta & _{i|i-1}^{(\alpha_{0:j-1},(v;\alpha^{(v)}),\alpha_{j+1:i})}(\ell)\nonumber \\
\negthinspace\negthinspace\negthinspace\negthinspace\negthinspace\negthinspace\negthinspace & \triangleq\begin{cases}
\bar{\Lambda}_{B,i}^{((v;\alpha^{(v)}),\alpha_{j+1:i})}(\ell), & \negthinspace\negthinspace\negthinspace\negthinspace s(\ell)=j\\
\bar{\Lambda}_{S,i|i-1}^{(\alpha_{s(\ell):j-1},(v;\alpha^{(v)}),\alpha_{j+1:i})}(\ell), & \negthinspace\negthinspace\negthinspace\negthinspace t(\ell)=j>s(\ell)\\
\bar{Q}_{S,i-1}^{(\alpha_{s(\ell):j-1})}(\ell), & \negthinspace\negthinspace\negthinspace\negthinspace t(\ell)=j-1\\
Q_{B,i}(\ell), & \negthinspace\negthinspace\negthinspace\negthinspace\ell\in\mathbb{B}_{j},\alpha^{(v)}=-1
\end{cases}\negthinspace,\label{eq:full_gibbs_sampling_sub_opt_indiv_ass_weight}
\end{align}
\begin{align*}
\Lambda & _{S,j+1:i}^{(\alpha_{j+1:i})}(x_{j:i},\ell)=\prod_{q=j+1}^{i}\Lambda_{S,q|q-1}^{(\alpha_{q})}(x_{q}|x_{q-1},\ell),\\
\bar{\Lambda} & _{B,i}^{((v;\alpha^{(v)}),\alpha_{j+1:i})}(\ell)\\
= & \int\negthinspace\tau_{0:j-1}^{(\alpha_{s(\ell):j-1})}(x_{s(\ell):j-1},\ell)P_{B,j}(\ell)f_{B,j}(x_{j},\ell)\\
\times & \,\,\psi_{j,Z_{j}^{(v)}}^{(v,\alpha^{(v)})}(x_{j},\ell)\Lambda_{S,j+1:i}^{(\alpha_{j+1:i})}(x_{j:i},\ell)dx_{s(\ell):i},\\
\bar{\Lambda} & _{S,i|i-1}^{(\alpha_{s(\ell):j-1},(v;\alpha^{(v)}),\alpha_{j+1:i})}(\ell)\\
= & \int\negthinspace\tau_{0:j-1}^{(\alpha_{s(\ell):j-1})}(x_{s(\ell):j-1},\ell)P_{S,j-1}(x_{j-1},\ell)\\
\times & \,f_{S,j|j-1}(x_{j}|x_{j-1},\ell)\psi_{j,Z_{j}^{(v)}}^{(v,\alpha^{(v)})}(x_{j},\ell)\Lambda_{S,j+1:i}^{(\alpha_{j+1:i})}(x_{j:i},\ell)dx_{s(\ell):i},
\end{align*}
$s(\ell)$ and $t(\ell)$ in (\ref{eq:full_gibbs_sampling_sub_opt_indiv_ass_weight})
are, respectively, the earliest and latest times on $\{j:i\}$ such
that $\ell$ exists. The right hand side of Eq. (\ref{eq:full_gibbs_sampling_sub_opt_indiv_ass_weight_dist})
can be interpreted as the (unnormalized) probability that (conditioned
on $\gamma_{\bar{j}}(\ell)$ and $Z_{1:k}$) label $\ell$ generates
measurement $z_{\alpha^{(v)}}^{(v)}$ at time $j$, abbreviated as
$\text{Pr}_{j}(\ell\sim z_{\alpha^{(v)}}^{(v)})$. 

The rationale for choosing (\ref{eq:full_gibbs_sampling_sub_opt_indiv_ass_weight_dist})
can be seen by substituting (\ref{eq:full_gibbs_sampling_sub_opt_indiv_ass_weight_dist})
into (\ref{eq:markov_weight_dist_full_gibbs}) and (\ref{eq:full_gibbs_sampling_min_markov_weight_dist-1}),
which gives
\[
\vartheta_{j}(\alpha|\gamma_{\bar{j}}(\ell),\ell)\propto\overset{V}{\underset{v=1}{\prod}}\overset{k\lor(t(\ell)+1)}{\underset{i=j}{\prod}}\negthinspace\negthinspace\negthinspace\negthinspace\negthinspace\eta_{i|i-1}^{(\gamma_{0:j-1}(\ell),(v;\alpha^{(v)}),\gamma_{j+1:i}(\ell))}(\ell),
\]
i.e., the product of probabilities $\text{Pr}_{j}(\ell_{n}\sim z_{\alpha^{(v)}}^{(v)})$,
$v=1:V$. Consequently, the approximation in (\ref{eq:full_gibbs_sampling_joint_multi_sensor_ass_dist})
boils down to approximating $\text{Pr}_{j}(\ell_{n}\sim z_{\alpha^{(1)}}^{(1)},...,z_{\alpha^{(V)}}^{(V)})$
by $\text{Pr}_{j}(\ell_{n}\sim z_{\alpha^{(1)}}^{(1)})\times...\times\text{Pr}_{j}(\ell_{n}\sim z_{\alpha^{(V)}}^{(V)}).$
This is reasonable, because the event space is $\alpha\in\{-1\}^{V}\uplus\Lambda_{j}^{(1:V)}$,
and conditioned on $Z_{1:k}$ and $\gamma_{\bar{j}}$, the probability
that $\ell$ generating a measurement from one sensor is almost independent
from generating a measurement from another sensor. Furthermore, similar
to \cite{Vo2019a}, the support of (\ref{eq:stationary_dist_gamma})
with the minimally-Markovian property contains the support of (\ref{eq:paramd_full_gibbs_weight})
as per Proposition \ref{prop:full_gibbs_sampling_minimally_markov_support}.
\begin{prop}
\label{prop:full_gibbs_sampling_minimally_markov_support} The support
of (\ref{eq:stationary_dist_gamma}) with minimally-Markovian property
contains the support of (\ref{eq:paramd_full_gibbs_weight}).
\end{prop}
Our proposed full Gibbs sampler (\textcolor{black}{Algorithm \ref{algo:full_gibbs}}),
uses the minimally-Markovian strategy above. Note that, once each
$\gamma'_{j}(\ell_{n})$ is sampled, the corresponding trajectory
posterior and the association weight are updated using (\ref{eq:paramd_multi_scan_multi_sensor_trajectory_update})
and (\ref{eq:paramd_multi_scan_multi_sensor_weight_update_1}). The
complexity of this algorithm is $\mathcal{O}(TV\sum_{j=1}^{k}P_{j}^{2}M_{j})$
(although it can be implemented in $\mathcal{O}(T\sum_{j=1}^{k}(P_{j}^{2}+P_{j}VM_{j}))$
by pre-calculating $\beta_{n}^{(k,v)}$ values), where $T$ denotes
the number of generated new samples, $M_{j}\triangleq\mathrm{max}_{v\in\{1:V\}}\{|Z_{j}^{(v)}|\}$
and $P_{j}\triangleq|\mathbb{B}_{j}\uplus\mathcal{L}(\gamma_{j-1})|$.
Indicatively, this complexity is $\mathcal{O}(kTVP^{2}\dot{M})$,
where $\dot{M}\triangleq\mathrm{max}{}_{j\in\{1:k\}}\{M_{j}\}$ and
$P\triangleq\mathrm{max}{}_{j\in\{1:k\}}\{P_{j}\}$. 

Similar to the single-sensor case \cite{Vo2019b}, the full Gibbs
sampler has an exponential convergence rate.
\begin{prop}
\label{prop:full_gibbs_sampler_convergence}Starting from any valid
initial state $\gamma_{1:k}\in\Gamma_{1:k}$, the Gibbs sampler described
by Algorithm \ref{algo:full_gibbs} converges to the stationary distribution
(\ref{eq:stationary_dist_gamma}) at an exponential rate. More specifically,
let $\pi^{j}$ denote the $j$-th power of the transition kernel.
Then, 
\[
\underset{\gamma_{1:k},\dot{\gamma}{}_{1:k}\in\Gamma_{1:k}}{\mathrm{max}}(\lvert\pi^{j}(\dot{\gamma}{}_{1:k}|\gamma_{1:k})-\pi(\dot{\gamma}{}_{1:k})\rvert)\leq(1-2\beta)^{\lfloor\frac{j}{h}\rfloor}
\]
where $h=k+1$, $\beta\triangleq\underset{\gamma_{1:k},\dot{\gamma}{}_{1:k}\in\Gamma_{1:k}}{\mathrm{min}}\pi^{h}(\dot{\gamma}{}_{1:k}|\gamma_{1:k})>0$
is the least likely $h$-step transition probability. 
\end{prop}
Algorithm \ref{algo:batch} provides the pseudocode for batch computation
of the multi-sensor GLMB posterior (\ref{eq:paramd_multi_scan_multi_sensor_posterior})
using the full Gibbs sampler (Algorithm \ref{algo:full_gibbs}). Note
that the factor sampler is used to initialize the full Gibbs sampler
with a set of $Q_{0}$ valid association histories, each of which
is used to generate a set of $T$ samples distributed according to
(\ref{eq:stationary_dist_gamma}). Since Algorithm \ref{algo:batch}
can be executed in parallel for each $q$, indicatively its complexity
is also $\mathcal{O}(kTVP^{2}\dot{M})$. Alternatively, the posterior
(\ref{eq:paramd_multi_scan_multi_sensor_posterior}) can be recursively
propagated, i.e., smoothing-while-filtering, as shown in Algorithm
\ref{algo:smoothing_while_filtering}. \textcolor{black}{Note that
Algorithms \ref{algo:batch} and \ref{algo:smoothing_while_filtering}
both produce the same posterior with the same complexity. However,
the smoothing-while-filtering algorithm allows us to obtain an estimate
of the multi-object trajectory at each point in time, while the batch
algorithm is an offline method, where the multi-object trajectory
can only be estimated after all the observations have been collected.
For a fixed complexity per time step, a moving window-based implementation
can be adopted, which incurs an $\mathcal{O}(LTVP^{2}\dot{M})$ complexity
per time step, where $L$ is the length of the window.} 
\begin{algorithm}[h]
\caption{Batch}
\label{algo:batch} \SetKwInOut{Input}{Input}\SetKwInOut{Output}{Output}

{\small{}\Input{$Q,R,T,Q_{0}$ (no. fact. samples),\\$H$ (no.
full samples)} \Output{$[G_{0:k}^{(h)}]_{h=1}^{H}$} \vspace{0.1cm}
}{\small\par}

\hrule

{\small{}\vspace{0.1cm}
}{\small\par}

{\small{}$[G_{0:k}^{(q)}]_{q=1}^{Q}:=\textsf{Factor-SampleJoint}(Q,R)$\;
$[G_{0:k}^{(q)}]_{q=1}^{Q_{0}}:=Q_{0}$ }\textcolor{black}{\small{}best
of}{\small{} $[G_{0:k}^{(q)}]_{q=1}^{Q}$\; \For{$q=1:Q_{0}$}{
$[G_{0:k}^{(q,t)}]_{t=1}^{\tilde{T}}:=\textsf{Unique}(\textsf{Full-Gibbs}(G_{0:k}^{(q)},T))$\;
} $[G_{0:k}^{(h)}]_{h=1}^{H}:=H$ }\textcolor{black}{\small{}best
of}{\small{} $\textsf{Unique}([G_{0:k}^{(q,t)}]_{(q,t)=(1,1)}^{(Q_{0},\tilde{T})})$\;
Normalize weights $[w_{0:k}^{(h)}]_{h=1}^{H}$\;}{\small\par}
\end{algorithm}
\begin{algorithm}[h]
\caption{Smoothing-while-Filtering}
\label{algo:smoothing_while_filtering} \SetKwInOut{Input}{Input}\SetKwInOut{Output}{Output}

{\small{}\Input{$[G_{0:k-1}^{(h)}]_{h=1}^{H_{k-1}},[T^{(h)}]_{h=1}^{H_{k-1}},T$}
\Output{$[G_{0:k}^{(h)}]_{h=1}^{H_{k}}$} \vspace{0.1cm}
}{\small\par}

\hrule

{\small{}\vspace{0.1cm}
}{\small\par}

{\small{}\For{$h=1:H_{k-1}$}{ $[G_{0:k}^{(h,t)}]_{t=1}^{T^{(h)}}:=$
$\textsf{Unique}(\textsf{Factor-Gibbs}(G_{0:k-1}^{(h)},T^{(h)}))$\;
} $[G_{0:k}^{(h)}]_{h=1}^{\bar{H}_{k}}:=\bar{H}_{k}$ }\textcolor{black}{\small{}best
of}{\small{} $[G_{0:k}^{(h,t)}]_{(h,t)=(1,1)}^{(H_{k-1},T^{(h)})}$\;
\For{$h=1:\bar{H}_{k}$}{ $[G_{0:k}^{(h,t)}]_{t=1}^{T}:=\textsf{Full-Gibbs}(G_{0:k}^{(h)},T)$\;
} $[G_{0:k}^{(h)}]_{h=1}^{H_{k}}:=H_{k}$ }\textcolor{black}{\small{}best
of}\textcolor{blue}{\small{} }{\small{}$\textsf{Unique}([G_{0:k}^{(h,t)}]_{h,t=(1,1)}^{(\bar{H}_{k},T)})$\;
Normalize weights $[w_{0:k}^{(h)}]_{h=1}^{H_{k}}$\;}{\small\par}
\end{algorithm}

\section{Numerical Experiments\label{sec:Num_Exps}}

This section evaluates the performance of the proposed (multi-scan)
multi-sensor GLMB smoother and benchmarks it against the multi-sensor
GLMB filter via simulations on single-sensor, two-sensors and four-sensor
scenarios.

Births, deaths and movements of an unknown and time-varying number
of objects are simulated in a 3D surveillance region over\textcolor{blue}{{}
${\color{black}k=100}{\color{blue}{\color{black}{\color{blue}}s}}$}.
The objects' 3D positions are measured using multiple sensors with
severely limited observability (detection probability) and high measurement
noise. There are $12$ births at\textcolor{blue}{{} ${\color{black}j=1s^{\mathrm{}}}$}\textcolor{black}{,}\textcolor{blue}{{}
${\color{black}20s^{\mathrm{}}}$}\textcolor{black}{,}\textcolor{blue}{{}
${\color{black}40s^{\mathrm{}}}$}\textcolor{black}{,}\textcolor{blue}{{}
${\color{black}60s^{\mathrm{}}}$ }\textcolor{black}{and}\textcolor{blue}{{}
${\color{black}80s^{\mathrm{}}}$} (respectively $3$, $3$, $2$,
$2$, and $2$ objects), and the probability
of survival is set at $P_{S}(x^{(\ell_{i})},\ell_{i})=0.95$. Two
of the objects born at time \textcolor{blue}{${\color{black}j=1s}$
}\textcolor{black}{die at}\textcolor{blue}{{} ${\color{black}70s^{\mathrm{}}}$}\textcolor{black}{,}
and the peak number of $10$ live objects occurs from\textcolor{blue}{{}
${\color{black}80s^{\mathrm{}}}$ }onwards. The three objects born
at\textcolor{blue}{{} ${\color{black}j=1s^{\mathrm{}}}$ }cross paths
at $[0,-400,0]^{T}$ at time\textcolor{blue}{{} ${\color{black}40s^{\mathrm{}}}$},
and two of the objects born at time\textcolor{blue}{{} ${\color{black}20s^{\mathrm{}}}$
}cross paths at $[300,-200,200]^{T}$ at time\textcolor{blue}{{} ${\color{black}59s^{\mathrm{}}}$},
constituting a more challenging multi-object tracking scenario. 

The kinematic state of an object is represented by a 6D state vector,
i.e., $x_{k}=[p_{x,k},\dot{p}_{x,k},p_{y,k},\dot{p}_{y,k},p_{z,k},\dot{p}_{z,k}]^{T}$
consisting of 3D position and velocity, and follows a constant velocity
motion model. The single object transition density is linear Gaussian
and given by $f_{S,k|k-1}(x_{k+1}^{(\ell)}|x_{k}^{(\ell)})=\mathcal{N}(x_{k+1}^{(\ell)};F_{k}x_{k}^{(\ell)},Q_{k})$
where
\[
F_{k}=I_{3}\otimes\left[\begin{array}{cc}
1 & \Delta\\
0 & 1
\end{array}\right],\hspace{0.8cm}Q_{k}=\sigma_{a}^{2}I_{3}\otimes\left[\begin{array}{cc}
\frac{\Delta^{4}}{4} & \frac{\Delta^{3}}{2}\\
\frac{\Delta^{3}}{2} & \Delta^{2}
\end{array}\right],
\]
$I_{3}$ is the $3\times3$ identity matrix, $\Delta=1s$ is the sampling
time, $\sigma_{a}=5\ m/s^{2}$, and $\otimes$ denotes the matrix
outer product. Births are modeled by a Labeled Multi-Bernoulli (LMB)
Process with (birth and spatial distribution) parameters $\{r_{B,k}(\ell_{i}),p_{B,k}^{(i)}(\ell_{i})\}_{i=1}^{4}$,
where $\ell_{i}=(k,i)\in\mathbb{B}_{k}$, $r_{B,k}(\ell_{i})=0.03$,
$p_{B,k}^{(i)}(x^{(\ell_{i})},\ell_{i})=\mathcal{N}(x^{(\ell_{i})};m_{B,k}^{(i)},Q_{B,k})$,
\begin{align*}
m_{B,k}^{(1)}=\  & (0.1,0,0.1,0,0.1,0)^{T},\\
m_{B,k}^{(2)}=\  & (400,0,-600,0,200,0)^{T},\\
m_{B,k}^{(3)}=\  & (-800,0,-200,0,-400,0)^{T},\\
m_{B,k}^{(4)}=\  & (-200,0,800,0,600,0)^{T},
\end{align*}
and $Q_{B,k}=\mathrm{diag}([10,10,10,10,10,10]^{2})$. 

The noisy 3D position of an object captured by sensor $v$ in the
surveillance region $[-1000,1000]\times[-1000,1000]\times[-1000,1000]m^{3}$,
takes the form $z_{k}^{(v)}=[z_{x,k}^{(v)},z_{y,k}^{(v)},z_{z,k}^{(v)}]^{T}$,
and modeled by the linear gaussian likelihood function $g_{k}^{(v)}(z_{k}^{(v)}|x_{k}^{(\ell)})=\mathcal{N}(z_{k}^{(v)};H_{k}^{(v)}x_{k}^{(\ell)},R_{k}^{(v)})$,
where $H_{k}^{(v)}=I_{3}\otimes\left[\begin{array}{cc}
1 & 0\end{array}\right]$, and $R_{k}^{(v)}=\mathrm{diag}([20,20,20]^{2})$. The probability
of detection of each sensor is $P_{D}^{(v)}(x^{(\ell_{i})},\ell_{i})=0.3$,
and clutter is modeled as a Poisson RFS with intensity $\kappa_{k}^{(v)}(z)=3.75\times10^{-10}m^{-3}$
over the surveillance region, i.e., a clutter rate of $3$ per scan.
This scenario is more challenging than those in \cite{Vo2019b,Vo2019a}
due to far lower detection probability and higher measurement noise
in a cluttered 3D environment. 

We plot single runs of the multi-sensor GLMB smoother and filter for
the same sets of measurements under the single-sensor, two-sensor
and four-sensor cases, and provide an analysis of the results in the
remainder of this section. 
\begin{figure*}
\includegraphics{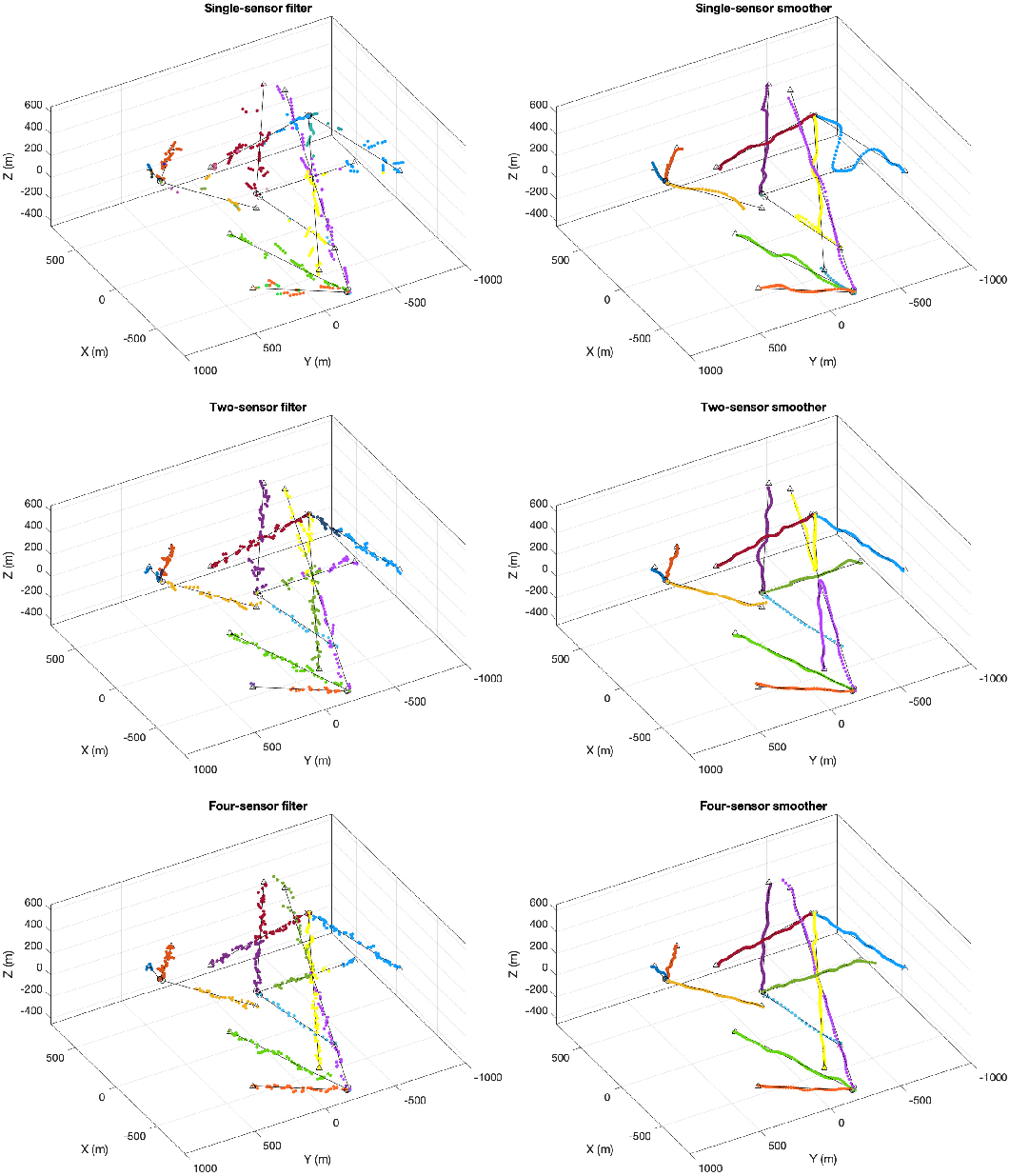}

\caption{\label{fig:Track_Comp}Estimated trajectories \textcolor{black}{(in
colored dots)} from GLMB filtering and GLMB smoothing superimposed
on true trajectories \textcolor{black}{(in black).} Starting and stopping
positions of objects are shown with $\Circle$ and $\ensuremath{\bigtriangleup}$
respectively.}

\end{figure*}

\begin{figure*}
\includegraphics{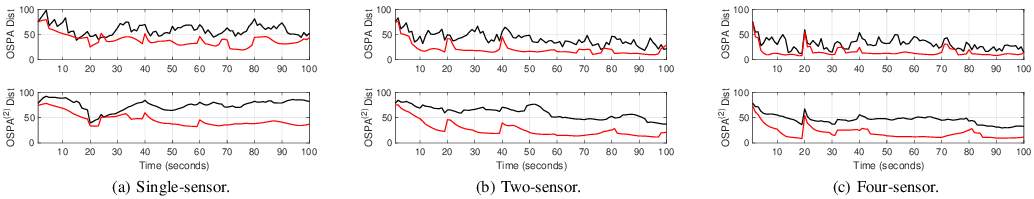}

\caption{\label{fig:OSPA_OSPA2_Comp}OSPA and OSPA$^{(2)}$ performance of
GLMB filtering (in black) and GLMB smoothing (in red).}

\vspace*{-0.4cm}
\end{figure*}

\vspace*{-0.4cm}

\subsection{Filtering vs. Smoothing}

The multi-sensor GLMB smoother (Algorithm \ref{algo:batch}) is run
with $100$ scans, and $1000$ valid initial samples generated via
factor sampling. Its performance is compared with the single-sensor
GLMB filter (Algorithm 2 of \cite{Vo2017}), single-sensor multi-scan
GLMB smoother (Algorithm 3 of \cite{Vo2019b} with the same parameters
as the proposed multi-sensor smoother), and multi-sensor GLMB filter
(Algorithm 3 of \cite{Vo2019a} using the minimally-Markovian strategy
with the same multi-sensor parameters). The smoothing problem involves
400 dimensional rank assignment problems with approximately 10 variables
in each dimension. State-of-the-art solvers with dedicated hardware
cannot handle problems with such large dimensions \cite{Reynen2019}.

Fig. \ref{fig:Track_Comp} shows the estimated trajectories from multi-sensor
GLMB filter and multi-sensor GLMB smoother (for single-sensor, two-sensor
and four-sensor scenarios) superimposed on the true trajectories.
Due to the challenging signal settings, the single-sensor GLMB filter
produces several fragmented tracks, fails to estimate large portions
of some tracks, and yields significant errors. Increasing the number
of sensors to two and four results in fewer fragmented tracks, produce
estimates for almost all portions of truth tracks, and improves estimates.
However, it does not prevent track switchings (at $[0,-400,0]^{T}$
and $[300,-200,200]^{T}$, respectively, for the two-sensor and four-sensor
scenarios) from occurring at track crossings.

The multi-sensor GLMB smoother outperforms the filter on each scenario.
The single-sensor scenario shows worse performance than the multiple
sensor scenarios due to lower observability, resulting in fragmented
tracks, undetected tracks and significant errors in two tracks. In
the two-sensor scenario, except for the two track switchings at $[0,-400,0]^{T}$,
all other tracks promptly start and terminate, resulting in smooth
trajectory estimates with smaller errors. The four-sensor scenario
produces superior results due to increased observability. It handles
the track crossings well, resulting in no track switchings, nor fragmented
tracks, and produces even smaller errors with each pertinent track
starting and terminating promptly. 

Fig. \ref{fig:OSPA_OSPA2_Comp} compares OSPA and OSPA$^{(2)}$ errors
over time between the multi-sensor GLMB smoother (final estimate)
and multi-sensor GLMB filter. The OSPA and OSPA$^{(2)}$ parameters
are $c=100m$, $p=1$, with a scan window size of $10$ for OSPA$^{(2)}$.
It is clear that increasing the number of sensors improves the performance
of the multi-sensor GLMB filter. The smoother produces significantly
lower errors due to its ability correct earlier errors in the multi-sensor
assignments.

\subsection{Posterior Statistics Computation}

\begin{figure*}
\includegraphics{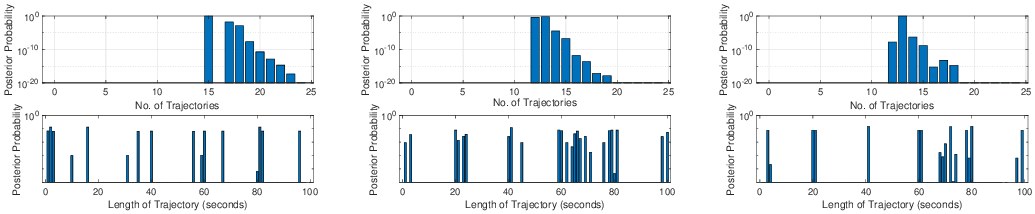}\caption{\label{fig:Res_NTraj_LTraj_Comp}Posterior distributions of the number
of trajectories, and posterior distributions of trajectory lengths.}
\end{figure*}

\begin{figure*}
\includegraphics{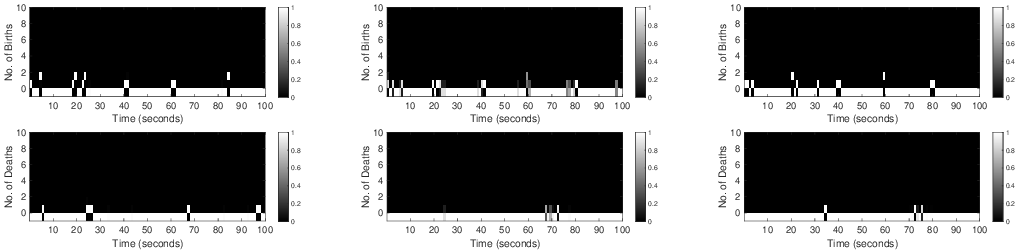}\caption{\label{fig:Res_Births_Deaths_Comp}Posterior distributions of the
births and deaths at each timestep.}
\end{figure*}

This section illustrates the proposed multi-object posterior capability
to provide useful statistical information \cite{Vo2019b} about the
ensemble of multi-object trajectories (\textcolor{black}{mathematical
expressions are given in eqs. (\ref{eq:ntraj_dist})-(\ref{eq:ltraj_dist})}).
Fig. \ref{fig:Res_NTraj_LTraj_Comp} shows the probability distribution
of the number of trajectories. As the number of sensors increases,
the mode of this distribution settles at $13$, although the actual
number of trajectories is $12$. This mismatch arises from a short
(\textcolor{blue}{${\color{black}3s}$}) false track appearing near
the birth location $(-800,-200,-400)^{T}$ due to a false measurement
falling close to it (recall that the detection probability of each
sensor is $0.3$). From the distribution of the length of trajectories
shown in Fig. \ref{fig:Res_NTraj_LTraj_Comp}, it can be seen that
except for the mode at time\textcolor{blue}{{} ${\color{black}j=3s}$}
(corresponding to the short false track), the posterior correctly
captures the modes at times\textcolor{blue}{{} }\textcolor{black}{$20s^{\mathrm{}}$,
$40s^{\mathrm{}}$, $60s^{\mathrm{}}$, $80s^{\mathrm{}}$ and $100s^{\mathrm{}}$
}in the four-sensor scenario. Also, note that the uncertainty at time\textcolor{blue}{{}
${\color{black}80s^{\mathrm{}}}$ }is higher in the two-sensor scenario
than the four-sensor scenario. Fig. \ref{fig:Res_Births_Deaths_Comp}
shows the birth and death cardinality distributions against time.
Except for the birth and death around time\textcolor{blue}{{} ${\color{black}30s^{\mathrm{}}}$
}(due to the short false track), the smoother correctly identifies
all instances of birth and deaths in the four-sensor scenario. 

\section{Conclusion\label{sec:Conclusion}}

In this paper, we developed an efficient numerical solution to the
multi-sensor multi-object smoothing problem via Gibbs sampling, which
unifies the implementation techniques for multi-sensor GLMB filtering
and multi-scan GLMB smoothing in \cite{Vo2019a,Vo2019b}. The proposed
solution, which involves solving large-scale NP-hard multi-dimensional
assignment problems, reduces to that of \cite{Vo2019b} when the number
of sensors is one and to that of \cite{Vo2019a} when the number of
scans is one. Theoretical justification and analysis of the proposed
algorithms are also presented. Numerical studies, conducted for the
first time with 100 scans and up to four sensors, show that excellent
tracking performance can be achieved despite having sensors with poor
detection probabilities (as low as 0.3) by combining them over a longer
integration time. This also demonstrates the benefits of multi-sensor
fusion for low observability sensors, which is vital for applications
in harsh signal environments such as underwater. 

\newpage{}

\end{document}